\documentclass[fleqn,usenatbib]{mnras}

\usepackage{newtxtext,newtxmath}

\usepackage[T1]{fontenc}

\DeclareRobustCommand{\VAN}[3]{#2}
\let\VANthebibliography\thebibliography
\def\thebibliography{\DeclareRobustCommand{\VAN}[3]{##3}\VANthebibliography}

\usepackage{graphicx}	
\usepackage{amsmath}	
\usepackage{amssymb}	
\usepackage{xcolor}
\usepackage{ulem}
\usepackage{censor}

\newcommand{\flux}{erg s$^{-1}$ cm$^{-2}$}
\newcommand{\sdss}{Sloan Digital Sky Survey}
\newcommand{\EBV}{$E(B-V)$}
\newcommand{\tauint}{$\tau_{\text{int}}$}
\newcommand{\mcmc}{Markov-Chain Monte Carlo}
\newcommand{\OIIIint}{the intrinsic [O III] $\lambda$5007}
\newcommand{\NIIint}{the intrinsic [N II] $\lambda$6584}
\newcommand{\ESS}{effective sample size}
\newcommand{\metal}{$12 + \text{log\,(O/H)}$}
\newcommand{\Ha}{H${\alpha}$}
\newcommand{\Hb}{H${\beta}$}
\newcommand{\Hg}{H${\gamma}$}

\newif\ifadd
\addfalse
\newif\ifdel
\delfalse
\newlength\nextcharwidth
\makeatletter
\renewcommand\@cenword[1]{%
  \setlength{\nextcharwidth}{\widthof{#1}}%
  \censorrule{\nextcharwidth}%
  \kern -\nextcharwidth%
  #1}
\makeatother
\newcommand{\add}[1]{\ifadd\textcolor{magenta}{#1}\else{#1}\fi}
\newcommand{\del}[1]{\ifdel\textcolor{red}{\sout{#1}}\else\fi}
\newcommand{\delref}[1]{\ifdel\textcolor{red}{\censorruledepth=.55ex\censor{#1}}\else\fi}

\title[Star-Forming Galaxies in HETDEX Pilot Survey]{Revisiting the Local Star-Forming Galaxies Observed in the HETDEX Pilot Survey}

\author[J.-H. Shinn]{
Jong-Ho Shinn,$^{1}$\thanks{E-mail: jhshinn@kasi.re.kr}
\\
$^{1}$Korea Astronomy and Space Science Institute, 776 Daeduk-daero, Yuseong-gu, Daejeon, 34055, the Republic of Korea
}

\date{Accepted XXX. Received YYY; in original form ZZZ}

\pubyear{2020}

\begin{document}
\label{firstpage}
\pagerange{\pageref{firstpage}--\pageref{lastpage}}
\maketitle

\begin{abstract}
    I have reanalyzed the data obtained for local ($z<0.15$) star-forming galaxies during the pilot survey for the Hobby-Eberly Telescope Dark Energy Experiment (HETDEX)---called the HETDEX Pilot Survey (HPS)---which uses an integral-field-unit spectrograph and covers $\sim3500-5800$ \AA{} at $\sim5$ \AA{} resolution.
    I have newly determined the gas metallicities, \metal, following the Bayesian analysis scheme of the previous study, but dealing carefully with the uncertainty of strong-line calibration, performing reproducibility tests with mock data, and monitoring the convergence of the \mcmc{} (MCMC) sampling.
    From the mock-data tests, I found that the nebular emission-line color excess \EBV{} can be overestimated by as much as 2-$\sigma$ or more, although the metallicity can recover the input value to within 1-$\sigma$.
    The new metallicity estimates on the HPS data are from well-converged MCMC samples (effective sample sizes $>$ 2000), and they are higher than the previous estimates by $\sim$2-$\sigma$.
    Using the HPS data, I also showed that the MCMC sampling can have the statistical accuracy as poor as the one near the iteration start if done without convergence monitoring.
    The overestimation of \EBV{} indicates the overestimation of the star-formation rates (SFRs) in the previous study, which can be as much as a factor of five.
    This finding undermines the previous suggestion of a hitherto-unknown galaxy population based on the locations of galaxies in the mass-SFR plane.
    I found that the independent determination of \EBV{} using either \Hb-\Hg{} or \Ha-\Hb{} line pair is ideal for the analysis of forthcoming HETDEX data, but it requires additional cost.
\end{abstract}




\section{Introduction} \label{intro}
Galaxies evolve after their formation through diverse processes such as star formation, stellar explosions, galaxy mergers, gas removal and accretion, etc.
These processes leave traces in various observable properties of the galaxies, one of which is metallicity \citep[see][]{Maiolino_2019_A&ARv_27_3}.
The metallicity $Z$ represents the mass of all metals (atoms heavier than helium) relative to the total mass of baryons (dominated by hydrogen and helium), but it is often expressed in terms of the oxygen abundance,
\begin{equation}
    12+\text{log\,(O/H)} \equiv 12+\text{log\,($N_O/N_H$)},
\end{equation}
where $N$ is the corresponding number density.
Since most of the metals originate in stellar interiors, the metallicity is closely related to the integrated amount of star formation in a galaxy over time.
Gas removal (e.g., outflows or stripping) and accretion of pristine gas also affect the metallicity.
The metallicity of stellar populations can be determined from stellar photospheric absorption lines, using model spectra from stellar population synthesis \citep[see][]{Conroy_2013_ARA&A_51_393,Maiolino_2019_A&ARv_27_3}.
Conversely, the metallicity of the gaseous components of a galaxy is mainly determined from emission lines originating in the interstellar medium, using the electron-temperature method, recombination lines, or a photoionization model \citep[see][]{Peimbert_2017_PASP_129_82001,Maiolino_2019_A&ARv_27_3}.
These three methods for gas metallicity estimation usually give discrepant results each other
\citep{Tsamis_2003_MNRAS_338_687,GarciaRojas_2007_ApJ_670_457,Kewley_2008_ApJ_681_1183,GarciaRojas_2009_A&A_496_139,Moustakas_2010_ApJS_190_233,LopezSanchez_2012_MNRAS_426_2630,GarciaRojas_2013_A&A_558_A122,Peimbert_2017_PASP_129_82001,ToribioSanCipriano_2017_MNRAS_467_3759}; the abundance difference can be as much as a factor of five between the electron-temperature and recombination lines methods \citep{Tsamis_2003_MNRAS_338_687}, and as much as $0.6-0.7$ dex between photoionization models and the other two \citep[e.g.,][]{Kewley_2008_ApJ_681_1183,Moustakas_2010_ApJS_190_233,LopezSanchez_2012_MNRAS_426_2630}.
Other than the three methods, the strong-line method is also used due to the weakness of the emission lines required to apply the electron-temperature or recombination lines methods \citep[see][]{Maiolino_2019_A&ARv_27_3}.
The strong-line method employs an empirically calibrated relation between the metallicity and the ratios of strong emission lines, and there exist diverse calibrations (see section \ref{ana-res-model}).

When one tries to extract any information---e.g., metallicity---from the observational data using a parameterized model, one must perform parameter estimation.
There are two distinct ways to do this: one is the frequentist approach (i.e., the classical approach), and the other is the Bayesian approach \citep[see][]{Wasserman_2004_book,Held_2013_book}.
The frequentist approach treats the true parameter(s) as unknown-but-fixed and the data as random.
In contrast, the Bayesian approach treats the true parameter(s) as random and the data as fixed.
Also, the frequentist approach uses the concept of confidence interval, while the Bayesian approach uses that of credible interval \citep[see][]{Held_2013_book}.
Bayesian parameter estimation is based on the posterior distribution, which is proportional to the product of the likelihood and the prior distribution\del{,} according to Bayes' theorem \citep[see][]{Sharma_2017_ARA&A_55_213}.
Many studies have adopted Bayesian parameter estimation because of its usefulness in inferring the information of interest from a given data set with the use of background knowledge \citep{Sharma_2017_ARA&A_55_213}.
This method has some pitfalls, however.
For instance, the posterior distribution can be much different from the likelihood due to the prior distribution, so reproducibility tests with mock data should be done after eliminating the effects of the prior distribution.
The posterior distribution is usually obtained using sampling methods such as \mcmc{} (MCMC, see \citealt{Sharma_2017_ARA&A_55_213}).
Since the MCMC method samples the target distribution based on random movements over a model parameter space, its convergence must be monitored to assess how close the sampled distribution is to the target distribution \citep{Hogg_2018_ApJS_236_11}; however, such convergence monitoring is often omitted in the literature.

Recently, \cite{Indahl_2019_ApJ_883_114} studied galactic gas metallicity using Bayesian parameter estimation.
More specifically, they reported gas metallicities and star formation rates (SFRs) for 29 low-redshift ($z<0.15$) galaxies, and they studied the distribution of those galaxies in mass-metallicity-SFR phase space.
Their targets are extraordinary, because they were selected using emission lines only---without any photometric (continuum-flux) preselection---over a wide area of the sky (\del{$\sim163$}\add{$\sim169$} arcmin$^2$).
In this way, they were able to explore for a new galaxy population that might have been missed in previous studies.
\cite{Indahl_2019_ApJ_883_114} determined the metallicities and the nebular emission-line color excesses \EBV{} using the strong-line method.
They followed the approach of \cite{GrasshornGebhardt_2016_ApJ_817_10}, and they carried out parameter estimation by employing the Bayesian approach and the MCMC method.
However, I noticed three points that \del{may undermine some of their conclusions}\add{can be improved}.
First, \cite{Indahl_2019_ApJ_883_114} adopted a relatively small uncertainty for the line-ratio calibration.
This affects the uncertainties of the estimated model parameters, since \cite{Indahl_2019_ApJ_883_114} included the scatter of the line ratio for a given metallicity as an uncertainty term in the likelihood.
Second, \delref{\cite{Indahl_2019_ApJ_883_114}}\del{ did not test the reproducibility of their methods with mock data.}\add{reproducibility test with mock data, which is not mentioned in \cite{Indahl_2019_ApJ_883_114}, will help determine how reliable the obtained results are.}
This test is important, because it can be hard to recover the true metallicity from the observed line ratios due to the scatter in the line-ratio calibration.
Third, \delref{\cite{Indahl_2019_ApJ_883_114}}\del{ did not monitor the convergence of their MCMC analysis.}\add{monitoring the convergence of MCMC sampling, which is also not mentioned in \cite{Indahl_2019_ApJ_883_114}, will enhance the statistical accuracy of the parameter estimates.}
As mentioned in the previous paragraph, convergence monitoring is essential for MCMC analyses, since it gives an estimate of the accuracy of the MCMC sampling results.
Also, \delref{\cite{Shinn_2019_MNRAS_489_4690}}\del{ reported that }we cannot secure the convergence by simply repeating the iterations\del{,} because of the recursive emergence of correlated samples\add{, as I reported in \cite{Shinn_2019_MNRAS_489_4690}}; hence again, convergence monitoring is important.

\del{Improper}\add{Improved} data-analysis leads to \del{improper}\add{improved} results and consequently \del{improper}\add{improved} conclusions.
Here I reanalyze the emission-line data of \cite{Indahl_2019_ApJ_883_114} following their analysis scheme but focusing on the three \del{issues}\add{points} mentioned above.
Then I see how \citeauthor{Indahl_2019_ApJ_883_114}'s results change and check which of \citeauthor{Indahl_2019_ApJ_883_114}'s conclusions \del{should be revised}\add{need to be reconsidered}.
I modeled the emission lines in the same way as \cite{Indahl_2019_ApJ_883_114}, but I adopted a larger scatter of the line ratio calibration in order to match the strong-line calibration that \cite{Indahl_2019_ApJ_883_114} selected, i.e., that of \cite{Maiolino_2008_A&A_488_463}.
I also performed several reproducibility tests with mock data, and I found that the metallicity is reproducible to within the 1-$\sigma$ level.
However, \EBV{} is poorly reproducible, and it can be overestimated by $>$ 2-$\sigma$; hence, the reddening-corrected SFRs of \cite{Indahl_2019_ApJ_883_114} are likely to be overestimated.
I also monitored the convergence during the MCMC sampling, and I found that the metallicity values of \cite{Indahl_2019_ApJ_883_114} are systematically lower than mine, mostly by 2-$\sigma$.

\section{Data Acquisition} \label{data}
Since I have reanalyzed the emission line data of \cite{Indahl_2019_ApJ_883_114}, all the data I used are the line fluxes reported in \cite{Indahl_2019_ApJ_883_114}.
Here I briefly describe how \cite{Indahl_2019_ApJ_883_114} obtained the spectra and line fluxes, and the reader is referred to \cite{Indahl_2019_ApJ_883_114} and \cite{Adams_2011_ApJS_192_5} for more information.

The 29 target galaxies in \cite{Indahl_2019_ApJ_883_114} are from the pilot survey for the Hobby-Eberly Telescope Dark-Energy Experiment (HETDEX, \citealt{Hill_2008_inbooka,Hill_2016_inproc}); hence, it is called the HETDEX Pilot Survey (HPS, \citealt{Adams_2011_ApJS_192_5}).
The HETDEX is a blind spectroscopic survey for a \del{420/4.5}\add{450} deg$^2$ area \add{(filling factor $\sim1/4.5$) }that is designed to provide a large sample of galaxies selected purely on the basis of their emission lines, without any photometric (continuum-flux) preselection.
The HETDEX uses an integral-field-unit (IFU) spectrograph called VIRUS \citep{Hill_2018_inproc} on the McDonald Observatory 10 m Hobby-Eberly Telescope; VIRUS has $\sim$35,000 1\farcs5 diameter fibers and covers $\sim3500-5500$ \AA{} with $\sim5.7$ \AA{} resolution.
The HPS is a HETDEX-like survey, but with a much smaller survey area (\del{$\sim163$}\add{$\sim169$} arcmin$^2$), and it was carried out using the George and Cynthia Mitchell Spectrograph (GMS, previously known as VIRUS-P: \citealt{Hill_2008_inbook}) on the McDonald Observatory 2.7 m Harlan J. Smith Telescope.
The GMS has 246 4\farcs2 diameter fibers, and it was configured to cover $\sim3500-5800$ \AA{} at $\sim5$ \AA{} resolution for the HPS (to be similar to the HETDEX).

\cite{Indahl_2019_ApJ_883_114} collected 29 low-redshift ($z<0.15$) galaxies that have [O II] $\lambda3727$, [O III] $\lambda5007$, or H$\beta$ lines from the HPS dataset.
In order to ensure that their galaxy list contains line-flux information for at least these three emission lines, \cite{Indahl_2019_ApJ_883_114} performed follow-up observations with another IFU spectrograph called LRS2 \citep{Chonis_2016_inbook} on the McDonald Observatory 10 m Hobby-Eberly Telescope.
The LRS2 has 280 fibers, each of which has a lenslet that covers a 0\farcs6 hexagonal field element, and it covers the wavelength range $\sim3700-10500$ \AA{} with a resolving power of $\sim1100-2000$.
\cite{Indahl_2019_ApJ_883_114} measured the line fluxes using a model consisting of Gaussian line(s) plus a linear continuum.
They tabulated a single summed flux for the line pair [O II] $\lambda\lambda3726,3729$ (reported as [O II] $\lambda3727$) and for the line pair [O III] $\lambda4959$, $\lambda5007$ (listed as [O III] $\lambda5007$).

\section{Analysis and Results}

\subsection{Line-Flux Model Test} \label{ana-res-model}
\cite{Indahl_2019_ApJ_883_114} estimated metallicities from the ratios of emission lines using the strong-line method, following the approach of \cite{GrasshornGebhardt_2016_ApJ_817_10}.
The strong-line method is a technique invented for an easier-but-less-precise estimation of metallicity.
The metallic emission lines used to determine the metallicity directly through the electron-temperature method or the recombination-line method are usually weaker than the Balmer lines by about a factor of $10-10^4$ \citep{Maiolino_2019_A&ARv_27_3}.
An alternative method was therefore developed to estimate the metallicity from strong emission lines, which can be detected more easily.
Calibration of the strong lines has been done empirically using the electron-temperature method \citep[e.g.,][]{Pettini_2004_MNRAS_348_L59,Pilyugin_2005_ApJ_631_231,Pilyugin_2010_ApJ_720_1738,Pilyugin_2016_MNRAS_457_3678,Curti_2017_MNRAS_465_1384}, a photoionization model \citep[e.g.,][]{Zaritsky_1994_ApJ_420_87,McGaugh_1991_ApJ_380_140,Kewley_2002_ApJS_142_35,Kobulnicky_2004_ApJ_617_240,Tremonti_2004_ApJ_613_898,Nagao_2011_A&A_526_A149,Dopita_2016_Ap&SS_361_61}, or both \citep[e.g.,][]{Denicolo_2002_MNRAS_330_69,Nagao_2006_A&A_459_85,Maiolino_2008_A&A_488_463}.
\cite{Indahl_2019_ApJ_883_114} adopted the calibration of \cite{Maiolino_2008_A&A_488_463}, which was done with local galaxies ($z\sim0$) using both the electron-temperature method and a photoionization model.
For the low-metallicity region [\metal{} $<8.3$], \cite{Maiolino_2008_A&A_488_463} used 259 galaxy samples from \cite{Nagao_2006_A&A_459_85}, for which metallicities were determined using the electron-temperature method.
For the high-metallicity region [\metal{} $>8.3$], they used 22,482 galaxies from the \sdss{} (SDSS, \citealt{York_2000_AJ_120_1579}) DR4 \citep{Adelman-McCarthy_2006_ApJS_162_38}, and they determined the metallicities with the photoionization model of \cite{Kewley_2002_ApJS_142_35}.

\cite{Indahl_2019_ApJ_883_114} modeled the emission-line flux using the strong-line calibration of \cite{Maiolino_2008_A&A_488_463}.
They used the following three line ratios:
\begin{equation} \label{R23}
    \text{R23}=\frac{\text{[O II]}\,\lambda3727 + \text{[O III]}\,\lambda4959 + \text{[O III]}\,\lambda5007}{\text{H}\beta},
\end{equation}
\begin{equation} \label{O32}
    \text{O32}=\frac{\text{[O III]}\,\lambda5007}{\text{[O II]}\,\lambda3727},
\end{equation}
\begin{equation} \label{N2}
    \text{N2}=\frac{\text{[N II]}\,\lambda6584}{\text{H}\alpha}.
\end{equation}
Using these three line ratios, I modeled the line fluxes with three parameters: the metallicity \metal{}, the intrinsic (i.e., not-reddened) [O III] $\lambda5007$ flux, and the intrinsic [N II] $\lambda6584$ flux\footnote{\cite{Indahl_2019_ApJ_883_114} used the intrinsic \Ha{} flux as a free parameter instead of \NIIint{} flux.}.
I fixed the ratio of [O III] $\lambda$5007 to [O III] $\lambda$4959 at 2.98 \citep{Storey_2000_MNRAS_312_813}, as in \cite{Indahl_2019_ApJ_883_114}.
A given metallicity, \metal{}, determines the ratios R23, O32, N2 and their uncertainties from the calibration function of \cite{Maiolino_2008_A&A_488_463}.
Then, \OIIIint{} flux determines the fluxes of other emission lines in the equations (\ref{R23}) and (\ref{O32}).
In a similar way, \NIIint{} determines the flux of the other emission line in the equation (\ref{N2}), i.e., \Ha.
The corresponding uncertainties of the line fluxes are calculated from eqs.~(\ref{R23})-(\ref{N2}) using the error propagation.

\begin{figure*}
    \center{
    \includegraphics[scale=0.65]{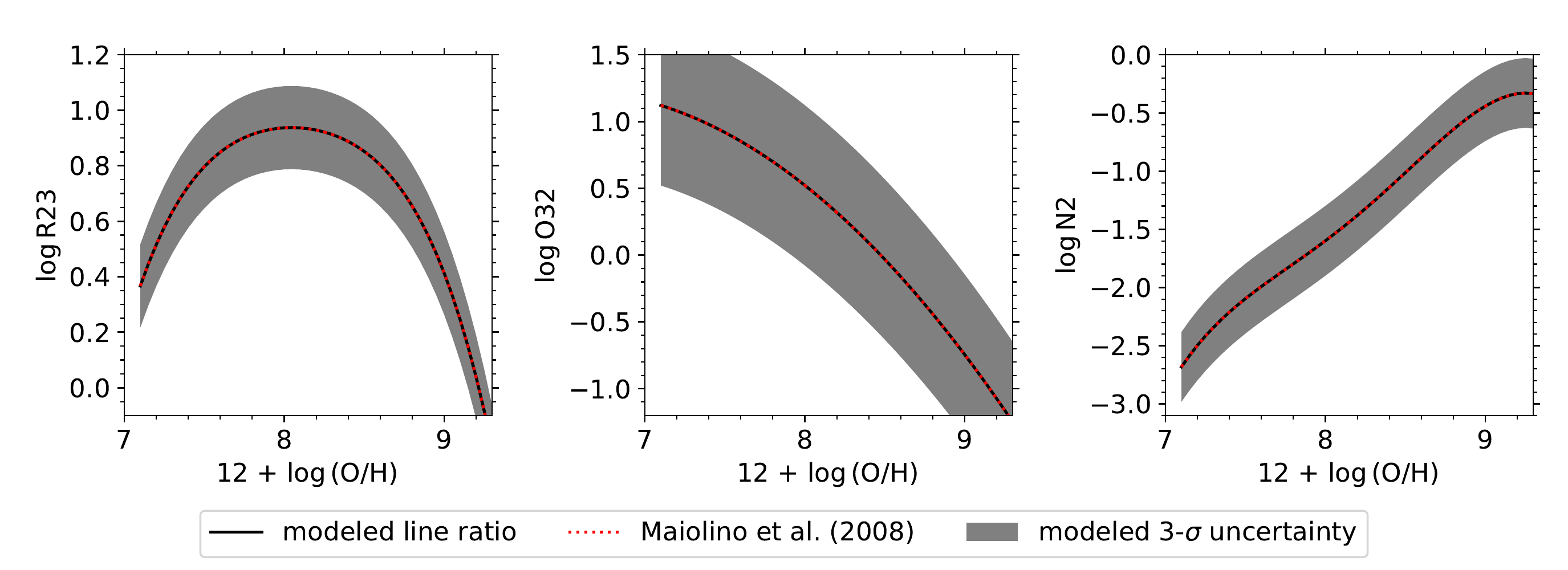}
    }
    \caption{Modeled line ratios and their uncertainties as a function of metallicity. The line ratios are defined in eqs. (\ref{R23})--(\ref{N2}). \protect\del{The upper panels show the model used in this study, while the lower panels show that of }\protect\delref{\protect\cite{Indahl_2019_ApJ_883_114}. }The red dotted line is the metallicity-calibration function derived by \protect\cite{Maiolino_2008_A&A_488_463}, and the black solid line is the line ratio calculated from the line fluxes modeled according to the metallicity-calibration function of \protect\cite{Maiolino_2008_A&A_488_463}. Gray shading indicates the modeled 3-$\sigma$ uncertainty region, which are calculated from the modeled uncertainties of the line fluxes.}
    \label{fig-model_test}
\end{figure*}

To check whether the line-flux modeling was done appropriately, I compared the line ratios calculated from the modeled line fluxes directly to the calibration of \cite{Maiolino_2008_A&A_488_463}.
Fig.~\ref{fig-model_test} shows a comparison of the line ratios.
The line ratios themselves clearly follow the calibration function of \cite{Maiolino_2008_A&A_488_463} very well.
\del{However, I found that the uncertainties adopted for the line ratios by }\delref{\cite{Indahl_2019_ApJ_883_114}}\del{ were too small for O32 and N2 (see the lower panel of Fig.~\ref{fig-model_test} of this work and Fig.~5 of }\delref{\citealt{Maiolino_2008_A&A_488_463}).}
\cite{Indahl_2019_ApJ_883_114} \add{had }adopted 10\% of each ratio as the uncertainty for all three line ratios\add{, but I found that it is small to mimic the spreads of the line ratios O32 and N2 (see Fig.~5 of \citealt{Maiolino_2008_A&A_488_463})}.
\del{Instead}\add{Therefore}, I adjusted the line-ratio uncertainty to cover most of the data points plotted in \cite{Maiolino_2008_A&A_488_463} at the 3-$\sigma$ level.
The values I adopted are as follows: $\Delta\,(\text{log\,O32})=0.2$, $\Delta\,(\text{log\,R23})=0.05$, and $\Delta\,(\text{log\,N2})=0.1$.
In addition, I set the correlation between ($\text{[O II]}\,\lambda3727 + \text{[O III]}\,\lambda4959 + \text{[O III]}\,\lambda5007$) and R23---which is arbitrary---to be $-1$, since this makes the modeled uncertainty most similar to the data-point scatter in the calibration plot of \cite{Maiolino_2008_A&A_488_463}.
\del{I here note that the uncertainty plotted in Fig.~6 of }\delref{\cite{Indahl_2019_ApJ_883_114}}\del{ is $2-3$ times larger than 10\% of the ratio, which they described they had adopted.
This may be a simple mistake in plotting, or it may be due to confusion between the natural logarithm and the base-10 logarithm.}

\subsection{Reproducibility Tests with Mock Data} \label{ana-res-mock}
To estimate the metallicity from the observed line fluxes using the line-flux model presented in section \ref{ana-res-model}, \cite{Indahl_2019_ApJ_883_114} took account of dust reddening using the Calzetti attenuation curve \citep{Calzetti_2000_ApJ_533_682} by adding one more parameter, \EBV{}, the nebular emission-line color excess.
They corrected the reddening in this way, because the HPS does not cover H$\alpha$ (see section \ref{data}) and hence they were unable to use the line ratio between H$\alpha$ and H$\beta$ for the reddening correction.
\cite{Indahl_2019_ApJ_883_114} then used the Bayesian approach and sampled the posterior distribution using the MCMC method.
Their log-likelihood expression has the form below:
\begin{equation} \label{likeli}
    \text{ln}\,\mathcal{L} \sim -\frac{1}{2}\sum_{l}^{} \frac{(x_{\text{obs},l}-x_{\text{mod},l})^2}{\sigma_{\text{obs},l}^2+\sigma_{\text{mod},l}^2}.
\end{equation}
Here $l$ means the different emission lines over which the fraction is summed; $x$ and $\sigma$ are the line flux and its uncertainty; and the subscripts `obs' and `mod' mean the corresponding values from the observations and the model, respectively.
This likelihood includes the uncertainty in the model line flux ($\sigma_{\text{mod},l}$), which can be large, since the scatter in the line ratios for a given metallicity is large \citep[see][]{Maiolino_2008_A&A_488_463}.
This can make it difficult to recover the true model values.
In this section, I test with mock data how well the likelihood is able to recover the model input values.
For this test, I first took flat priors to see the effects of likelihood only.
The prior ranges are as follows: \metal{}, (6.5, 10.0); \EBV{}, (0, 0.79); \OIIIint{} flux, (0, $10^{-8}$) \flux; and \NIIint{} flux, (0, $10^{-8}$) \flux.
The metallicity range is the same with the one of \cite{Indahl_2019_ApJ_883_114}, which is a little wider than the calibration range of \cite{Maiolino_2008_A&A_488_463}, (7.0, 9.3).
The \EBV{} range is based on the prior used by \cite{Indahl_2019_ApJ_883_114}: a Gaussian prior with $\sigma=0.165$ centered at 0.295.
\cite{Indahl_2019_ApJ_883_114} obtained these two values from local SDSS star-forming galaxies.
For the flat prior for \EBV{}, I set the maximum to be the Gaussian mean + 3 $\times$ Gaussian sigma.
For the intrinsic line fluxes, I used a sufficiently large range.

\begin{figure*}
    \center{
    \includegraphics[scale=0.65]{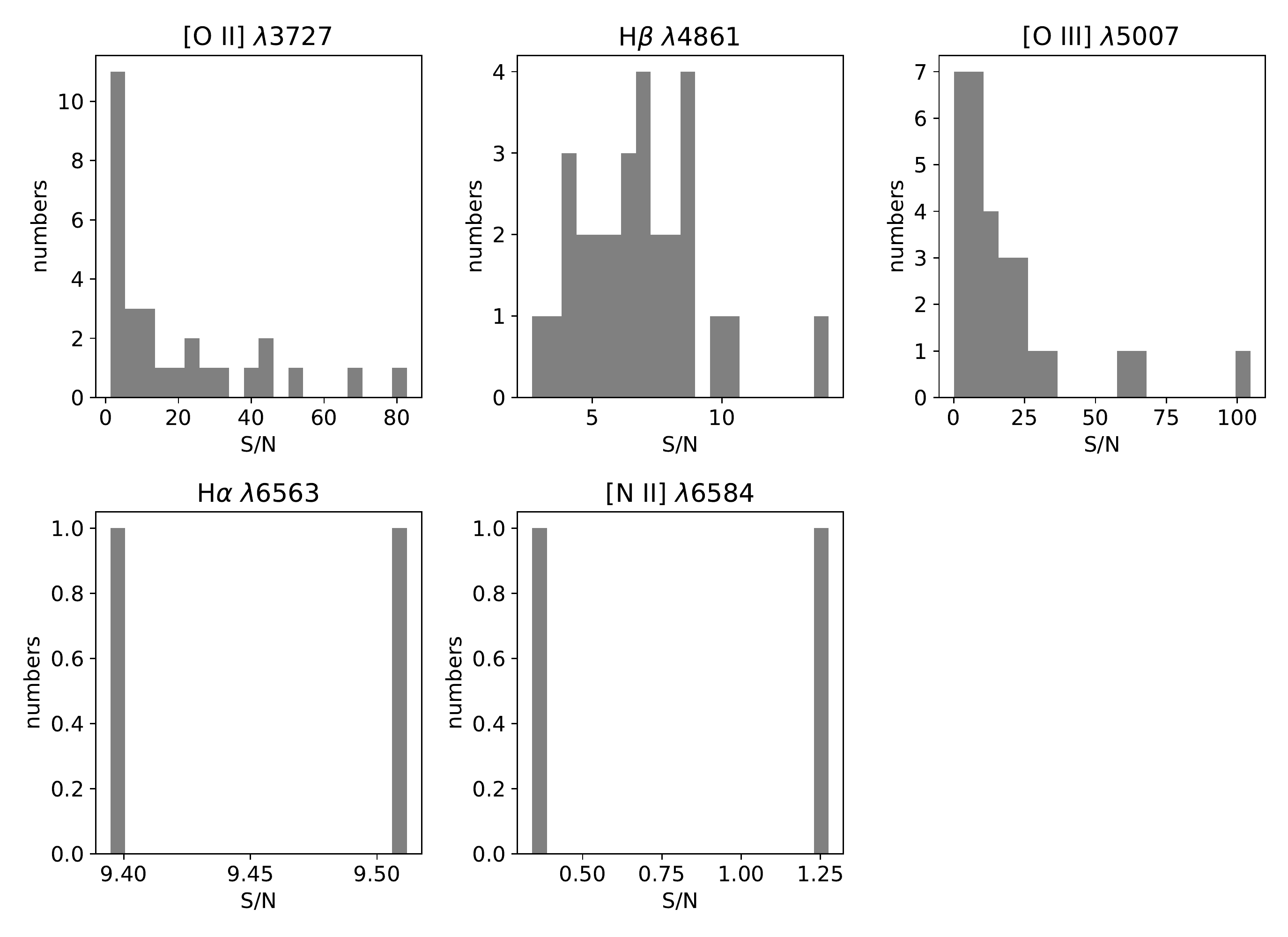}
    }
    \caption{Signal-to-noise ratio (S/N) distributions for all the emission lines reported in \protect\cite{Indahl_2019_ApJ_883_114}. [O II] $\lambda$3727 represents the sum of the [O II] $\lambda$3726 and $\lambda$3729 lines, while [O III] $\lambda$5007 represents the sum of the [O III] $\lambda$4959 and $\lambda$5007 lines (see \protect\citealt{Indahl_2019_ApJ_883_114}).}
    \label{fig-snr}
\end{figure*}

First, I created mock data with four model input values: \metal{}, \EBV{}, \OIIIint{} flux, and \NIIint{} flux.
To set the line-flux uncertainty for the mock data, I referred to the signal-to-noise ratio (S/N) distribution for the observational data tabulated in \cite{Indahl_2019_ApJ_883_114}.
Fig.~\ref{fig-snr} shows the S/N distribution of all the observed emission lines.
The S/N ratio can be as high as $\sim100$, but it is mostly $\la10$.
Thus, I started the test with mock data having S/N = 100, excluding the model uncertainty from the likelihood.
Second, I carried out MCMC sampling of the posterior distribution, employing the affine-invariant ensemble sampler called \textsf{emcee} \citep{Foreman-Mackey_2013_PASP_125_306,Foreman-Mackey_2019_JOSS_4_1864}.
I used the stretch move \citep{Goodman_2010_CAMCoS_5_65} with the stretch scale parameter $a=2$.
For MCMC initialization I used the mode of the posterior distribution \citep[see][]{Hogg_2018_ApJS_236_11,Shinn_2019_MNRAS_489_4690}.
To find it, I employed a global optimization method called differential evolution \citep{Storn_1997_J.GlobalOptim._11_341}.
I repeated the optimization process 32 times and used the 32 results for the MCMC initialization; hence, the  number of walkers for the MCMC sampling is also 32.

\begin{figure*}
    \center{
    \includegraphics[scale=0.65]{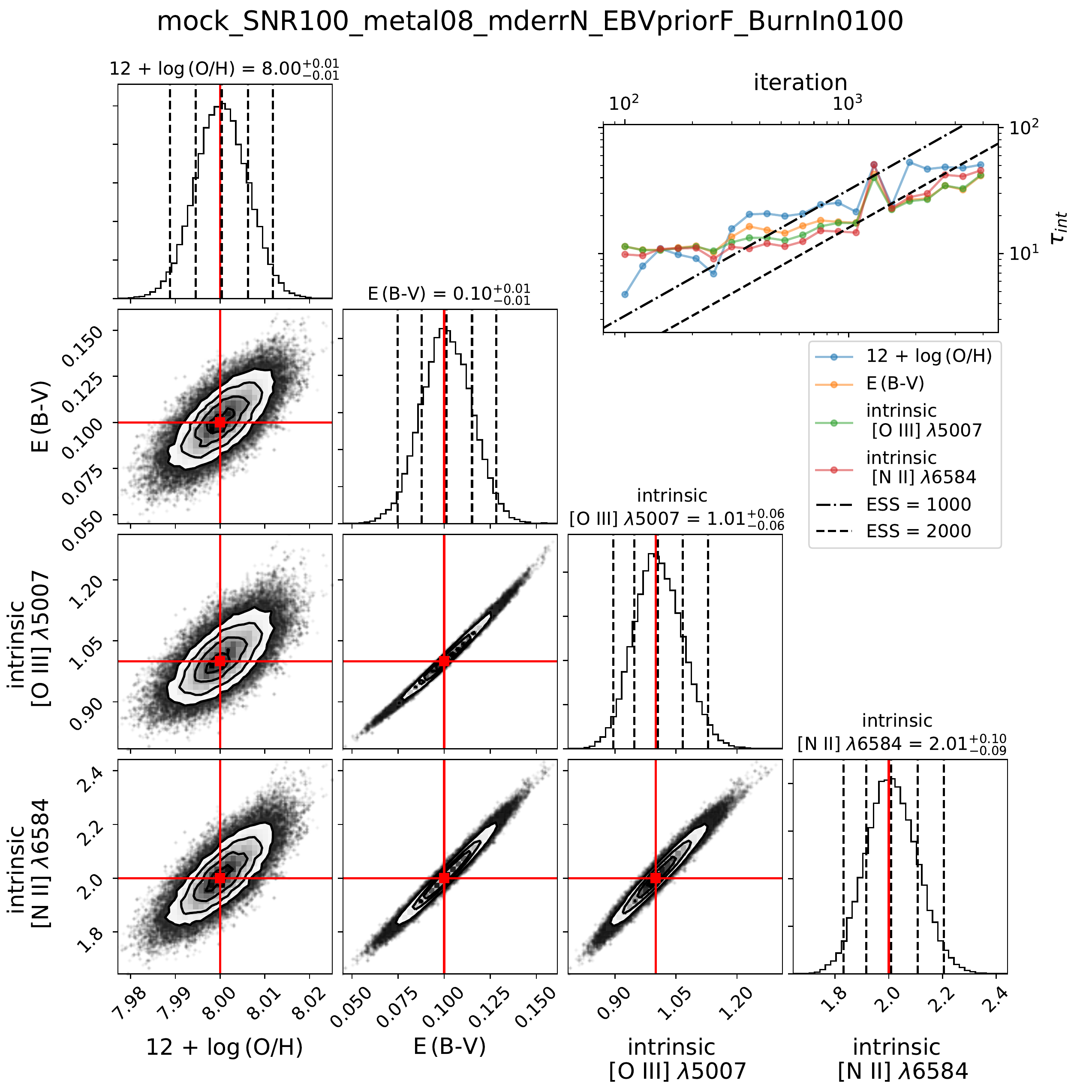}
    }
    \caption{Posterior distribution and evolution of the integrated autocorrelation times (\tauint) from a mock-data test. The ten panels in the lower left corner are the corner plot of the Markov-Chain Monte Carlo (MCMC) sampling results. It shows the correlations among the model parameters and their marginal distributions. Two names `intrinsic [O III] $\lambda5007$' and `intrinsic [N II] $\lambda6584$' indicate \del{`\OIIIint{} flux' and `\NIIint{} flux', respectively}\add{the corresponding line fluxes in units of $10^{-16}$ \flux}. The red solid lines are the input values used to generate the mock data. The vertical dashed lines indicate the median, 1-$\sigma$ (68 \%), and 2-$\sigma$ (95 \%) credible intervals. The panel in the upper right corner shows the evolution of \tauint. For convergence diagnosis, I also plot two straight lines, which correspond to effective sample sizes (ESSs) of 1000 and 2000, respectively. The title at the top of the figure indicates the following: SNR\#\#\# (signal-to-noise ratio of the mock data), metal08 [metallicity \metal{} = 8.0], mderrN (model uncertainty excluded), EBVpriorF [flat prior for \EBV{}], and BurnIn\#\#\#\# (burn-in iterations excluded before plotting).}
    \label{fig-mock-snr100}
\end{figure*}

Fig.~\ref{fig-mock-snr100} shows the posterior distribution for the mock data with S/N = 100, where the model uncertainty is excluded from the likelihood.
This figure shows that the model input values are well reproduced, and the model parameters all have positive correlations with each other.
The tight positive correlations among \EBV{}, \OIIIint{}\footnote{The word `flux' is dropped from `\OIIIint{} flux' for simplicity from here on.}, and \NIIint{}\footnote{The word `flux' is dropped from `\NIIint{} flux' for simplicity from here on.} are reasonable, because a higher \EBV{} requires higher line fluxes for a fixed metallicity.
In the upper-right corner of Fig.~\ref{fig-mock-snr100}, I plot the evolution of the integrated autocorrelation time (\tauint{}) to monitor the convergence of the MCMC sampling.
MCMC sampling has only N/\tauint{} independent samples, where N is the total sample length \citep{Sharma_2017_ARA&A_55_213}, because the samples from dynamic Monte Carlo methods are usually correlated \citep{Sokal_2013_inbook}.
The number N/\tauint{} is called the effective sample size (ESS), and I also plot two ESS lines as convergence checks in the upper-right panel in Fig.~\ref{fig-mock-snr100}.
The ESS lines are equal to N/\tauint{} multiplied by the number of walkers (i.e., 32), since I calculated \tauint{} from the mean of the ensemble walkers (i.e., $X_i=\frac{1}{32}\sum^{32}_{i=1} x_i$), as in \cite{Goodman_2010_CAMCoS_5_65}.
I used the following definition of \tauint{}:
\begin{equation} \label{eq-iat}
    \tau_{\text{int}}=\sum^{\infty}_{t=-\infty}\rho_{xx}(t)
    \text{, where } \rho_{xx}(t)=\frac{\mathbb{E}[(x_i-\bar{x})(x_{i+t}-\bar{x})]}{\mathbb{E}[(x_i-\bar{x})^2]}.
\end{equation}
Here $\rho_{xx}$ is the autocorrelation function for the sequence $\{x_i\}$, $t$ is the time difference---or distance---between two points in the sequence $\{x_i\}$, $\bar{x}$ is the mean of sequence $\{x_i\}$, and $\mathbb{E}\left[\cdot\right]$ means the expectation value.
I calculated $\tau_{int}$ with a routine in the \textsf{emcee} package, using an ``automatic windowing'' size of 5 \citep[see][]{Sokal_2013_inbook}.
I considered the sampling to be sufficiently converged when the values of \tauint{} for all of the parameters cross the ESS = 2000 line (Fig.~\ref{fig-mock-snr100}).
This value of ESS is larger than the number 1665 that is required to determine the 0.025 quantile to within 0.0075 with probability 0.95, which corresponds to about a 10\% error in the 0.025 quantile for light-tailed (normal) or moderate-tailed (student's $t_{\nu=4}$) distributions \citep{Raftery_1992_inproc}.
After achieving convergence, I checked to determine whether there exist any unexplored local extrema by increasing the stretch-scale parameter by a factor of five (i.e., $a=10$) in the MCMC sampler, as in \cite{Shinn_2019_MNRAS_489_4690}.
I obtained nil for this check and it is the same for all the following tests shown in this section.

\begin{figure*}
    \center{
    \includegraphics[scale=0.65]{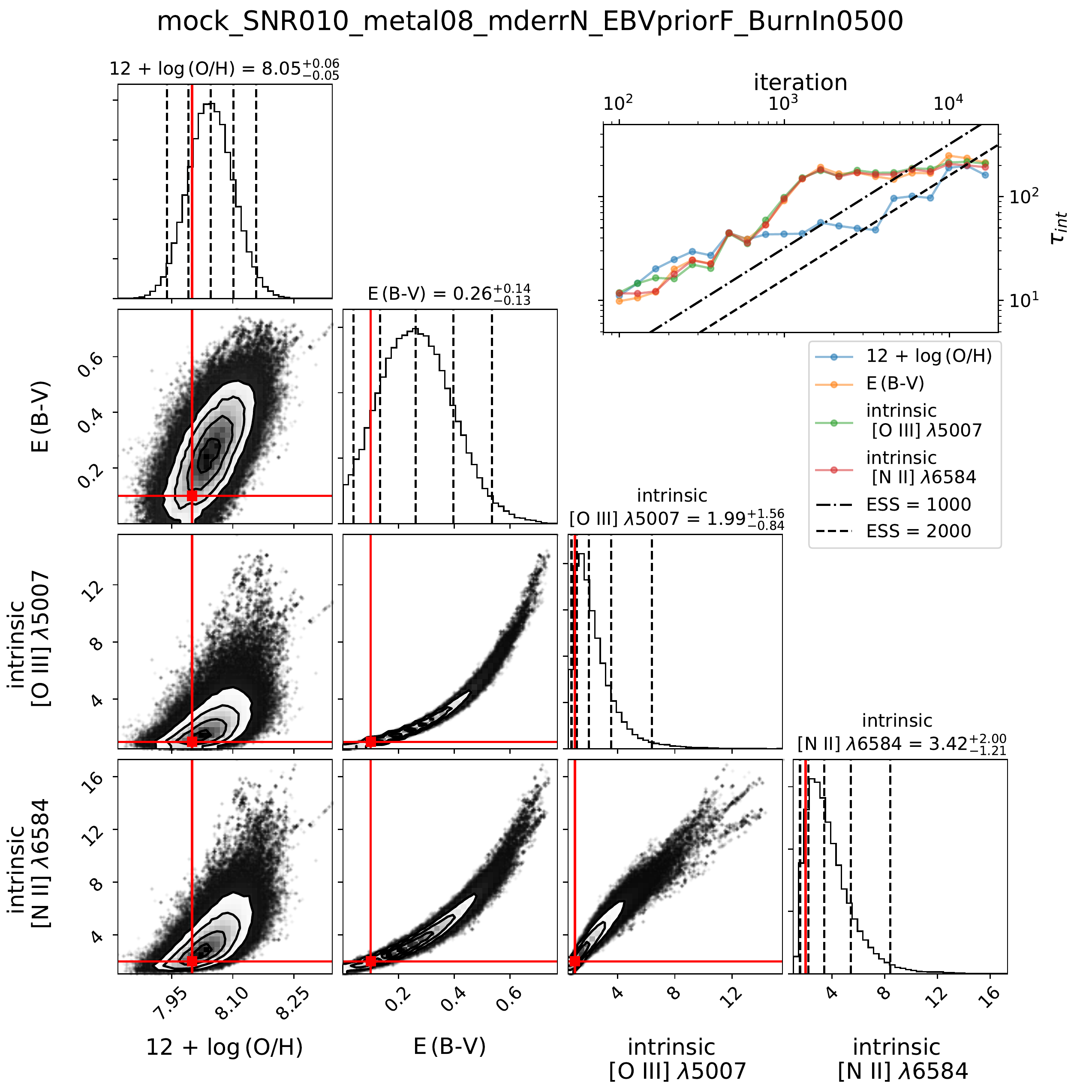}
    }
    \caption{Posterior distribution and evolution of the integrated autocorrelation times (\tauint) from another mock-data test. The setting for this test is the same as for Fig.~\ref{fig-mock-snr100}, except that I changed the signal-to-noise ratio of the mock data from 100 to 10 (and hence changed the title to SNR010). The figure description is otherwise the same as for Fig.~\ref{fig-mock-snr100}.}
    \label{fig-mock-snr10}
\end{figure*}

Next, I reduced the S/N of the mock data to 10, which is a bit higher than the most probable S/N of the observed line fluxes (Fig.~\ref{fig-snr}), and examined how the posterior distribution changes.
Fig.~\ref{fig-mock-snr10} shows the results.
As the marginal distributions show, all the parameters are overestimated, but they recover the input model values around the 1-$\sigma$ level.
Obviously, the lowered S/N of 10 causes the overestimation (compare Fig.~\ref{fig-mock-snr100} and \ref{fig-mock-snr10}).
In other words, the higher $\sigma_{\text{obs},l}$ in eq.~(\ref{likeli}) deforms the likelihood and consequently the posterior.
Note that the uncertainty of the model line flux ($\sigma_{\text{mod},l}$) is not included in the likelihood [eq.~(\ref{likeli})] under this test.
The \EBV{} and \OIIIint{} pair, as well as the \EBV{} and \NIIint{} pair, start to show narrow and curved covariances (Fig.~\ref{fig-mock-snr10}).

\begin{figure*}
    \center{
    \includegraphics[scale=0.65]{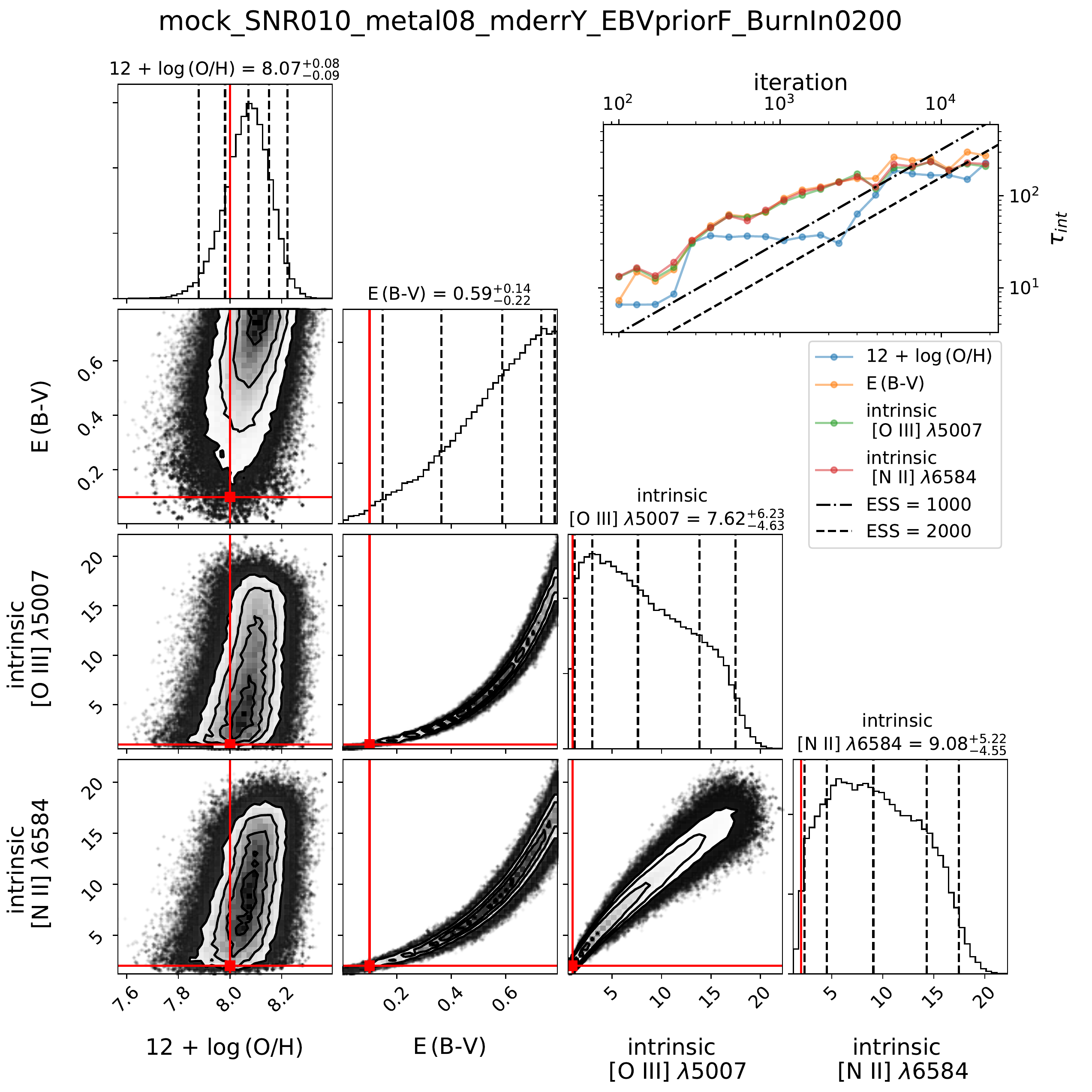}
    }
    \caption{Posterior distribution and evolution of the integrated autocorrelation times (\tauint) from another mock-data test. The setting for this test is the same as for Fig.~\ref{fig-mock-snr10}, except that I have included the model uncertainty (hence the title mderrY). The figure description is otherwise the same as for Fig.~\ref{fig-mock-snr100}.}
    \label{fig-mock-mderrY}
\end{figure*}

As a third test, I included the model uncertainties ($\sigma_{\text{mod},l}$) into the likelihood [eq.~(\ref{likeli})], and Fig.~\ref{fig-mock-mderrY} shows the results.
The prominent changes in the posterior distribution are that the three model parameters---\EBV{}, \OIIIint{}, and \NIIint{}--- suffer severely from overestimation.
The input values of the mock data fall well below the medians of the posterior parameter distributions at the 2-$\sigma$ level or more.
This drastic overestimation for the three parameters is due to the inclusion of the model uncertainties, which stem from the uncertainty of the strong-line calibration (see Fig.~\ref{fig-model_test}), into the likelihood [eq.~(\ref{likeli})].
However, note that the posterior distribution of the metallicity recovers the input value of the mock data well at the 1-$\sigma$ level, although it is still an overestimate.

\begin{figure*}
    \center{
    \includegraphics[scale=0.65]{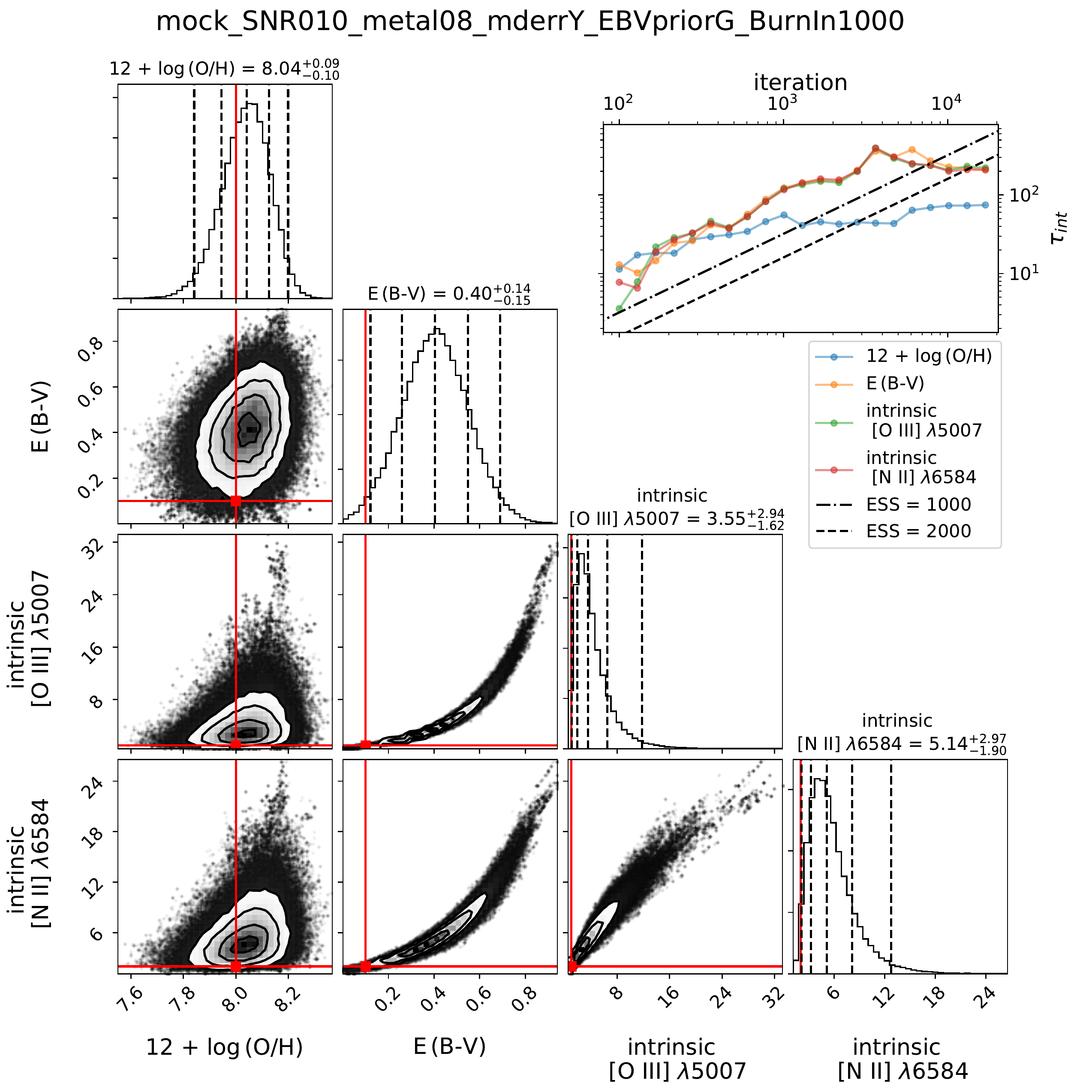}
    }
    \caption{Posterior distribution and evolution of the integrated autocorrelation times (\tauint) from another mock-data test. The setting for this test is the same as for Fig.~\ref{fig-mock-mderrY}, except that I adopted a Gaussian prior for \EBV{} (hence the title EBVpriorG). The figure description is otherwise the same as for Fig.~\ref{fig-mock-snr100}.}
    \label{fig-mock-EBVpriorG}
\end{figure*}

\begin{figure*}
    \center{
    \includegraphics[scale=0.65]{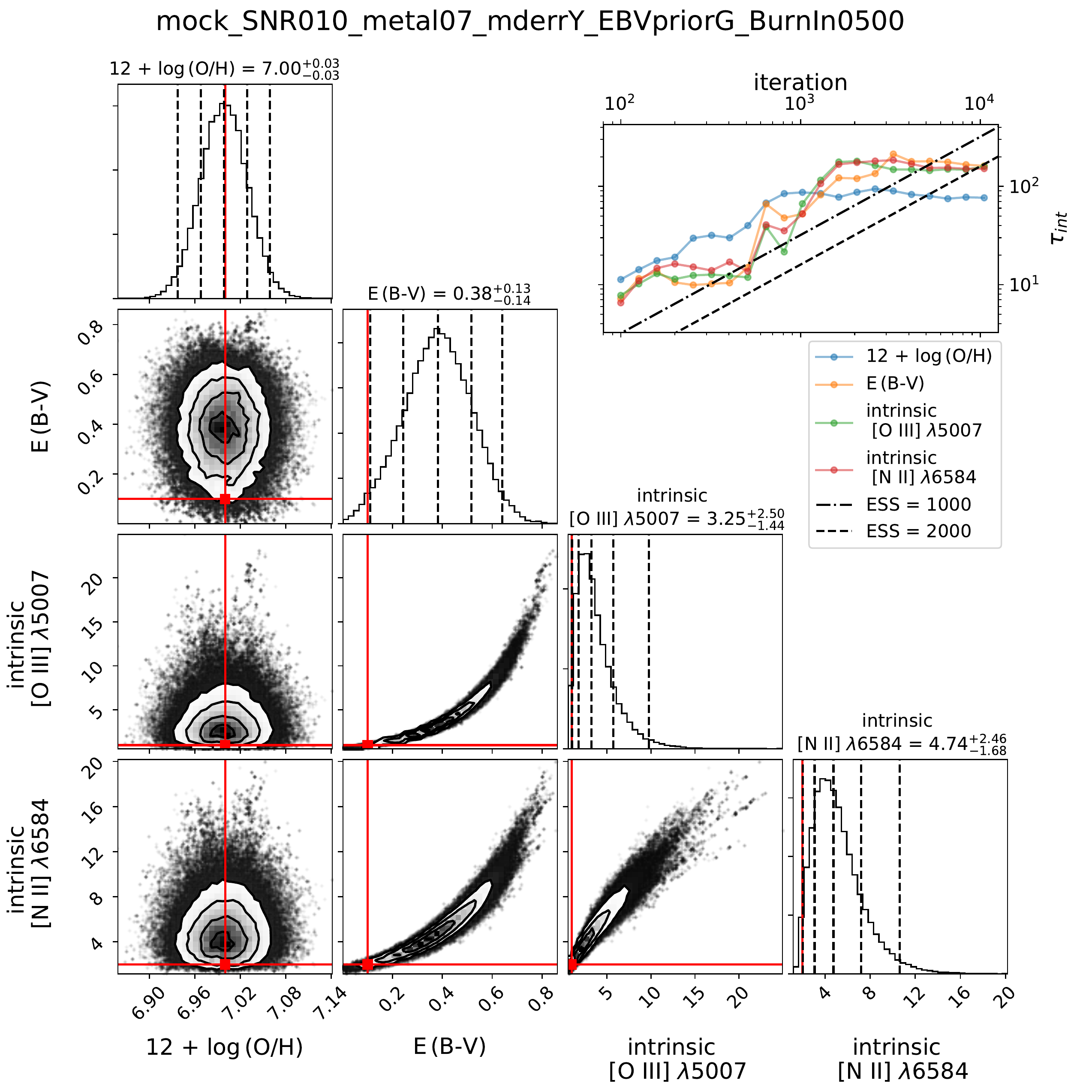}
    }
    \caption{Posterior distribution and evolution of the integrated autocorrelation times (\tauint) from another mock-data test. The setting for this test is the same as for Fig.~\ref{fig-mock-EBVpriorG}, except that I changed the metallicity from 8 to 7 (hence the title metal07). The figure description is otherwise the same as for Fig.~\ref{fig-mock-snr100}.}
    \label{fig-mock-metal7}
\end{figure*}

\begin{figure*}
    \center{
    \includegraphics[scale=0.65]{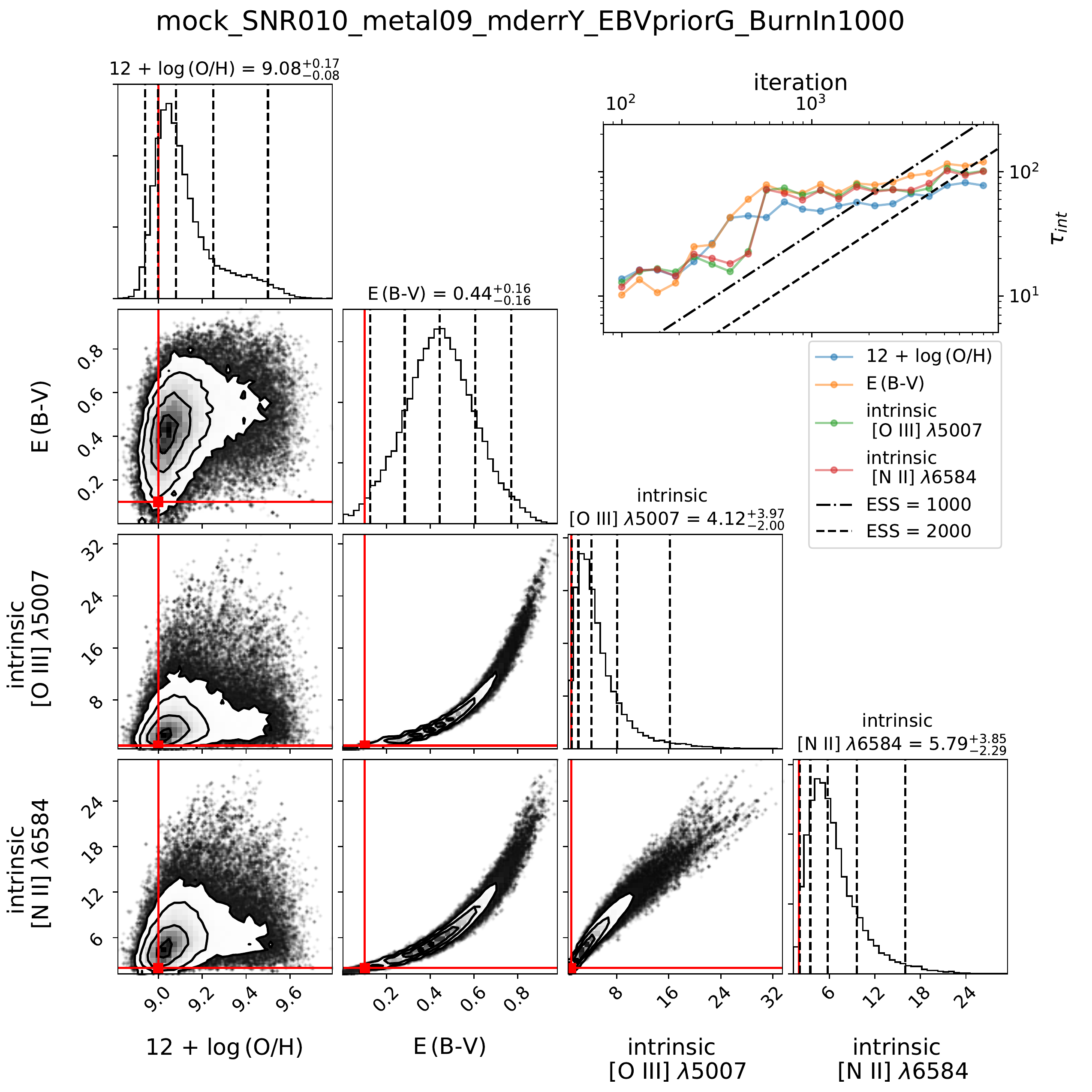}
    }
    \caption{Posterior distribution and evolution of the integrated autocorrelation times (\tauint) from another mock-data test. The setting for this test is the same as for Fig.~\ref{fig-mock-metal7}, except that I have changed the metallicity from 7 to 9 (hence the title metal09). The figure description is otherwise the same as for Fig.~\ref{fig-mock-snr100}.}
    \label{fig-mock-metal9}
\end{figure*}

As a final test, I replaced the flat prior for \EBV{} with the one used by \cite{Indahl_2019_ApJ_883_114}: a Gaussian prior with $\sigma=0.165$ centered at 0.295.
Fig.~\ref{fig-mock-EBVpriorG} shows the results.
As the marginal distributions show, the overall trend is still the same as in the previous test (Fig.~\ref{fig-mock-mderrY}). 
The three parameters---\EBV{}, \OIIIint{}, and \NIIint{}---are still overestimated at the 2-$\sigma$ level or more, while the metallicity is well recovered within the 1-$\sigma$ level.
I tested two more cases by setting \metal{} = 7.0 (Fig.~\ref{fig-mock-metal7}) and \metal{} = 9.0 (Fig.~\ref{fig-mock-metal9}) as the input values for the mock data, but the overestimation trends remained.
For comparison, I also ran an additional test with a higher input value of \EBV{} = 0.3 (near the mode of the prior distribution), but the overestimation trends still remained, except that the overestimates of the three parameters are reduced to 1-$\sigma$.
In the case of the input \EBV{} = 0.5, all four parameters were well recovered within 1-$\sigma$.

From all the test results shown in this section, I draw the following two conclusions:
First, the metallicity can be recovered to within the 1-$\sigma$ level by the method of \cite{Indahl_2019_ApJ_883_114}.
Second, the remaining three parameters---\EBV{}, \OIIIint{}, and \NIIint{}---can be overestimated by as much as 2-$\sigma$ or more by the method of \cite{Indahl_2019_ApJ_883_114}.

\subsection{Reanalysis of the Data of \citeauthor{Indahl_2019_ApJ_883_114}} \label{ana-res-hps}
I reanalyzed the line fluxes observed by \cite{Indahl_2019_ApJ_883_114} as done above for the mock data.
One exception was that I reduced the metallicity prior range from (6.5, 10.0) to (7.0, 9.3) to make it equal to the metallicity calibration range of \cite{Maiolino_2008_A&A_488_463}.
I initialized the MCMC sampler at the mode of the posterior distribution, and I sampled that distribution until the MCMC samples converged sufficiently---i.e., \tauint{} for each of the parameters crossed the line ESS = 2000.
Then I checked to determine whether there are any unexplored local extrema by increasing the stretch scale parameter by a factor of five ($a=10$), and I obtained nil for all the analyzed galaxies.
Here I only analyze 27 [O II]-selected galaxies and ignore the two [O III]-selected galaxies \citep[see Table 3 and 4 of][]{Indahl_2019_ApJ_883_114}; hence, only the ratios R23 and O32 were used.
There are two reasons for this.
First, \cite{Indahl_2019_ApJ_883_114} did not provide the H$\alpha$ absorption values, so I could not correct the observed H$\alpha$ line flux, which is needed to use the ratio N2 [eq.(\ref{N2})].
Second, for the [O III]-selected galaxies, \cite{Indahl_2019_ApJ_883_114} did not tabulate the model parameter values estimated by using the ratios R23 and O32 only (excluding N2), so I could not compare my model parameter values to theirs.
As a result, only three parameters---\metal{}, \EBV, and \OIIIint---were used for the modeling; I did not need \NIIint{} parameter since the [O II]-selected galaxies do not have the \Ha{} and [N II] $\lambda$6584 flux data.

\begin{figure*}
    \center{
    \includegraphics[scale=0.65]{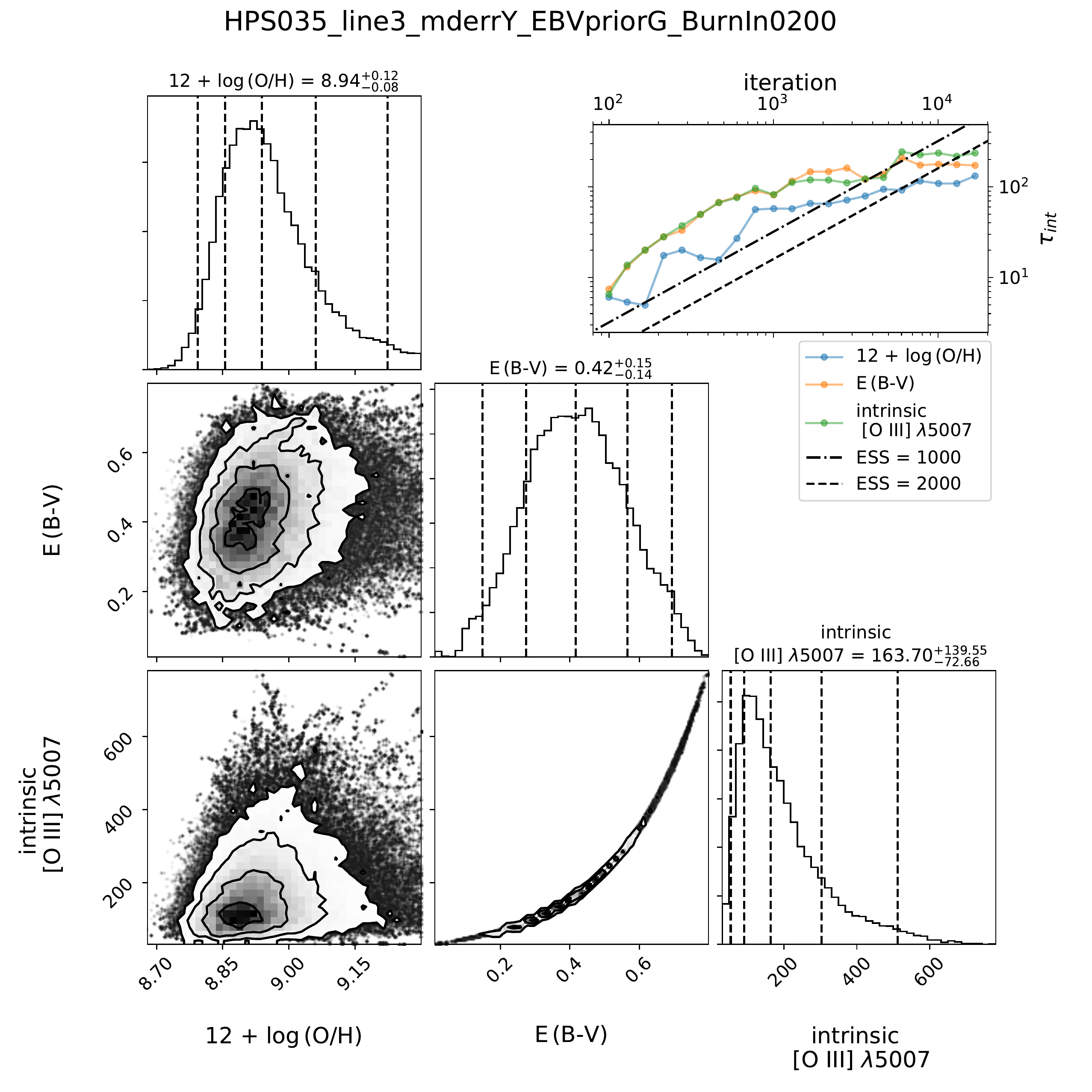}
    }
    \caption{Posterior distribution and evolution of the integrated autocorrelation times (\tauint) for HPS035. The complete figure set of 27 galaxies is available in the online supplementary data. The six panels in the lower left corner are the corner plot of the Markov-Chain Monte Carlo (MCMC) sampling results, which shows the correlations among the model parameters and their marginal distributions.\add{{ }The name `intrinsic [O III] $\lambda5007$' indicates the corresponding line flux in units of $10^{-16}$ \flux.{ }}The vertical dashed lines indicate the median, 1-$\sigma$ (68 \%), and 2-$\sigma$ (95 \%) credible intervals. The panel in the upper right corner shows the evolution of \tauint. For convergence diagnosis, I also plot two straight lines corresponding to the effective sample sizes (ESSs) of 1000 and 2000, respectively. The title at the top of the figure indicates the following: HPS\#\#\# (HPS ID), line3 (number of data points used for the analysis), mderrY (model uncertainty included), EBVpriorG [Gaussian prior for \EBV{}], and BurnIn\#\#\#\# (burn-in iterations excluded before plotting).}
    \label{fig-hps}
\end{figure*}

\begin{table}
	\centering
	\caption{Model-parameter estimates from the \mcmc{} (MCMC) analysis. Column (1) is the HPS ID; Column (2) is the metallicity; Column (3) is the nebular emission-line color excess; and Column (4) is \OIIIint{} flux. The values in Columns (3) and (4) are not reliable because of overestimation (see section \ref{ana-res-mock}).}
	\label{tbl-mcmc}
	\begin{tabular}{cccc} 
		\hline
        {} & {} & {} & {intrinsic}\\
        {HPS ID} & {12+log(O/H)} & {\EBV{}} & {[O III] $\lambda5007$\add{{ }flux}}\\
        & & & {($10^{-16}$ \flux{})}\\
        {(1)} &  {(2)} & {(3)} & {(4)}\\
		\hline
        HPS035 & ${8.94}^{+0.12}_{-0.08}$ & ${0.42}^{+0.15}_{-0.14}$ & ${163.70}^{+139.55}_{-72.66}$\\
HPS044 & ${8.84}^{+0.14}_{-0.10}$ & ${0.39}^{+0.15}_{-0.15}$ & ${11.28}^{+10.16}_{-5.33}$\\
HPS065 & ${9.06}^{+0.12}_{-0.09}$ & ${0.37}^{+0.15}_{-0.15}$ & ${23.00}^{+21.66}_{-11.08}$\\
HPS067 & ${9.12}^{+0.11}_{-0.12}$ & ${0.40}^{+0.17}_{-0.16}$ & ${3.76}^{+5.50}_{-2.35}$\\
HPS105 & ${8.77}^{+0.15}_{-0.10}$ & ${0.40}^{+0.15}_{-0.16}$ & ${23.22}^{+20.84}_{-11.07}$\\
HPS118 & ${8.95}^{+0.15}_{-0.11}$ & ${0.38}^{+0.16}_{-0.16}$ & ${5.94}^{+5.82}_{-2.92}$\\
HPS119 & ${8.88}^{+0.13}_{-0.09}$ & ${0.39}^{+0.15}_{-0.15}$ & ${51.48}^{+44.94}_{-23.86}$\\
HPS125 & ${8.92}^{+0.12}_{-0.09}$ & ${0.42}^{+0.16}_{-0.16}$ & ${63.15}^{+58.70}_{-30.51}$\\
HPS129 & ${8.72}^{+0.14}_{-0.10}$ & ${0.42}^{+0.15}_{-0.15}$ & ${32.95}^{+29.59}_{-15.66}$\\
HPS138 & ${8.93}^{+0.13}_{-0.08}$ & ${0.40}^{+0.15}_{-0.16}$ & ${20.30}^{+17.43}_{-9.70}$\\
HPS158 & ${8.90}^{+0.15}_{-0.10}$ & ${0.38}^{+0.15}_{-0.15}$ & ${8.25}^{+7.47}_{-3.90}$\\
HPS219 & ${8.98}^{+0.14}_{-0.10}$ & ${0.37}^{+0.16}_{-0.16}$ & ${24.42}^{+22.66}_{-11.66}$\\
HPS225 & ${8.90}^{+0.13}_{-0.09}$ & ${0.39}^{+0.15}_{-0.15}$ & ${11.93}^{+10.62}_{-5.47}$\\
HPS235 & ${8.92}^{+0.18}_{-0.17}$ & ${0.39}^{+0.16}_{-0.16}$ & ${5.23}^{+5.93}_{-2.86}$\\
HPS237 & ${9.01}^{+0.14}_{-0.13}$ & ${0.43}^{+0.16}_{-0.16}$ & ${5.15}^{+6.61}_{-3.10}$\\
HPS260 & ${8.89}^{+0.15}_{-0.12}$ & ${0.39}^{+0.15}_{-0.15}$ & ${4.63}^{+4.29}_{-2.24}$\\
HPS278 & ${8.70}^{+0.15}_{-0.10}$ & ${0.40}^{+0.16}_{-0.16}$ & ${58.89}^{+54.38}_{-28.53}$\\
HPS300 & ${8.99}^{+0.12}_{-0.09}$ & ${0.40}^{+0.16}_{-0.15}$ & ${6.97}^{+6.39}_{-3.32}$\\
HPS303 & ${8.77}^{+0.14}_{-0.11}$ & ${0.43}^{+0.16}_{-0.16}$ & ${16.47}^{+15.83}_{-8.09}$\\
HPS326 & ${8.94}^{+0.13}_{-0.09}$ & ${0.38}^{+0.15}_{-0.16}$ & ${13.34}^{+11.94}_{-6.48}$\\
HPS363 & ${9.04}^{+0.15}_{-0.13}$ & ${0.36}^{+0.17}_{-0.15}$ & ${5.46}^{+5.94}_{-2.77}$\\
HPS375 & ${8.97}^{+0.12}_{-0.08}$ & ${0.39}^{+0.14}_{-0.15}$ & ${71.38}^{+57.99}_{-33.19}$\\
HPS386 & ${8.94}^{+0.13}_{-0.09}$ & ${0.39}^{+0.16}_{-0.15}$ & ${26.84}^{+25.36}_{-12.61}$\\
HPS413 & ${9.01}^{+0.12}_{-0.09}$ & ${0.37}^{+0.15}_{-0.15}$ & ${22.74}^{+20.10}_{-10.51}$\\
HPS438 & ${8.96}^{+0.14}_{-0.10}$ & ${0.38}^{+0.15}_{-0.15}$ & ${4.61}^{+4.21}_{-2.23}$\\
HPS449 & ${8.84}^{+0.14}_{-0.09}$ & ${0.40}^{+0.16}_{-0.16}$ & ${18.67}^{+16.87}_{-8.84}$\\
HPS458 & ${9.07}^{+0.11}_{-0.09}$ & ${0.36}^{+0.15}_{-0.15}$ & ${10.29}^{+9.24}_{-4.78}$\\

		\hline
	\end{tabular}
\end{table}

Fig.~\ref{fig-hps} shows the posterior distributions obtained from the MCMC sampling results, and Table \ref{tbl-mcmc} lists the parameter values determined from the marginal posterior distributions.
I show only the HPS035 result here in Fig.~\ref{fig-hps}, but the complete results for 27 galaxies are available in the online supplementary data.
Overall, the posterior distributions show covariance shapes similar to those obtained for the mock data whose input metallicity is 8.0 (Fig.~\ref{fig-mock-EBVpriorG}) and 9.0 (Fig.~\ref{fig-mock-metal9}).
This is reasonable considering that all the metallicity estimates fall between $\sim$8.5 and $\sim$9.0 (Table \ref{tbl-mcmc}).
Three cases (HPS067, HPS235, and HPS237) have a bit different covariance shapes from the rest, and it seems that the low S/Ns of [O II] and [O III] emission lines cause it; these three targets have the three lowest S/Ns in [O II] and [O III] fluxes \citep[Table 3 of][]{Indahl_2019_ApJ_883_114}.

\begin{figure*}
    \center{
    \includegraphics[scale=0.65]{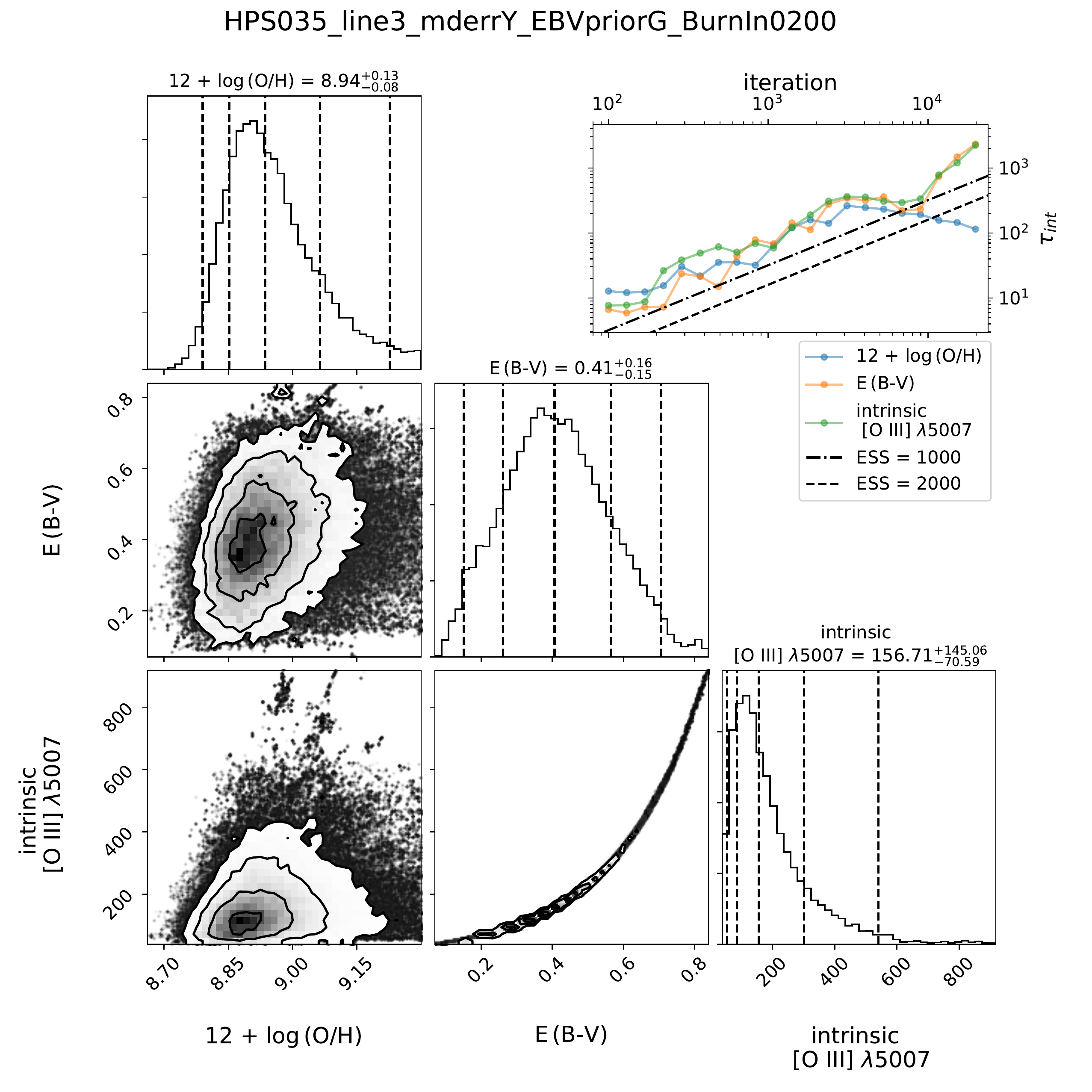}
    }
    \caption{Posterior distribution and evolution of the integrated autocorrelation times (\tauint) for HPS035, but from a single run of 20,000 iterations (without monitoring the convergence) instead of from multiple step-by-step runs (while monitoring the convergence, as shown in Fig.~\ref{fig-hps}). The upper right panel shows that \tauint{} increases abruptly at the end of the iteration for both \EBV{} and \OIIIint{}, making the \ESS{} (ESS) fall below 1000 again. The figure description is otherwise the same as for Fig.~\ref{fig-hps}.}
    \label{fig-hps-single}
\end{figure*}

Here I note that I had to run the MCMC sampler for each galaxy on a step-by-step basis to secure convergence by monitoring the evolution of \tauint{}.
In other words, I had to run the sampler using multiple short iterations, because \tauint{} soars abruptly at some point due to the emergence of correlated samples; I reported this phenomenon previously in \cite{Shinn_2019_MNRAS_489_4690}.
When \tauint{} soars abruptly at a certain run, I truncated that part and reran the sampler again at the end of the remaining sample.
When I ran the MCMC sampler mindlessly for a long iteration instead, \tauint{} usually continued to increase, and hence the ESS did not increase.
Fig.~\ref{fig-hps-single} shows a single long-iteration example for the target HPS035 (compare it to Fig.~\ref{fig-hps}).
At the end of the iteration (iteration $\sim10,000$), \tauint{} for \EBV{} and for \OIIIint{} abruptly increase, making the corresponding ESSs fall below 1000.
The statistical accuracy of the posterior distributions for these two parameters at the end of the iteration is therefore no better than at the start of the iteration, say at iteration = 100.
In other words, the posterior distribution from the $32\times20,000$ samples has the statistical accuracy as poor as the one from the $32\times100$ samples (Fig.~\ref{fig-hps-single}); the number 32 is the number of walkers used for the MCMC sampling (see section \ref{ana-res-mock}).
Note that the median and 1-$\sigma$ intervals of the marginal distributions are similar between the two cases (Figs.~\ref{fig-hps} and \ref{fig-hps-single}), but there is no comparison between them in terms of statistical accuracy.
This clearly shows how important it is to monitor the convergence with \tauint{} and ESS when carrying out MCMC sampling\footnote{An example of how to run MCMC sampling with the convergence monitoring is posted on \del{\url{http://vo.kasi.re.kr/Stat_Reanal/}}\add{\url{https://data.kasi.re.kr/vo/Stat_Reanal/}}}.

\begin{figure*}
    \center{
    \includegraphics[scale=0.65]{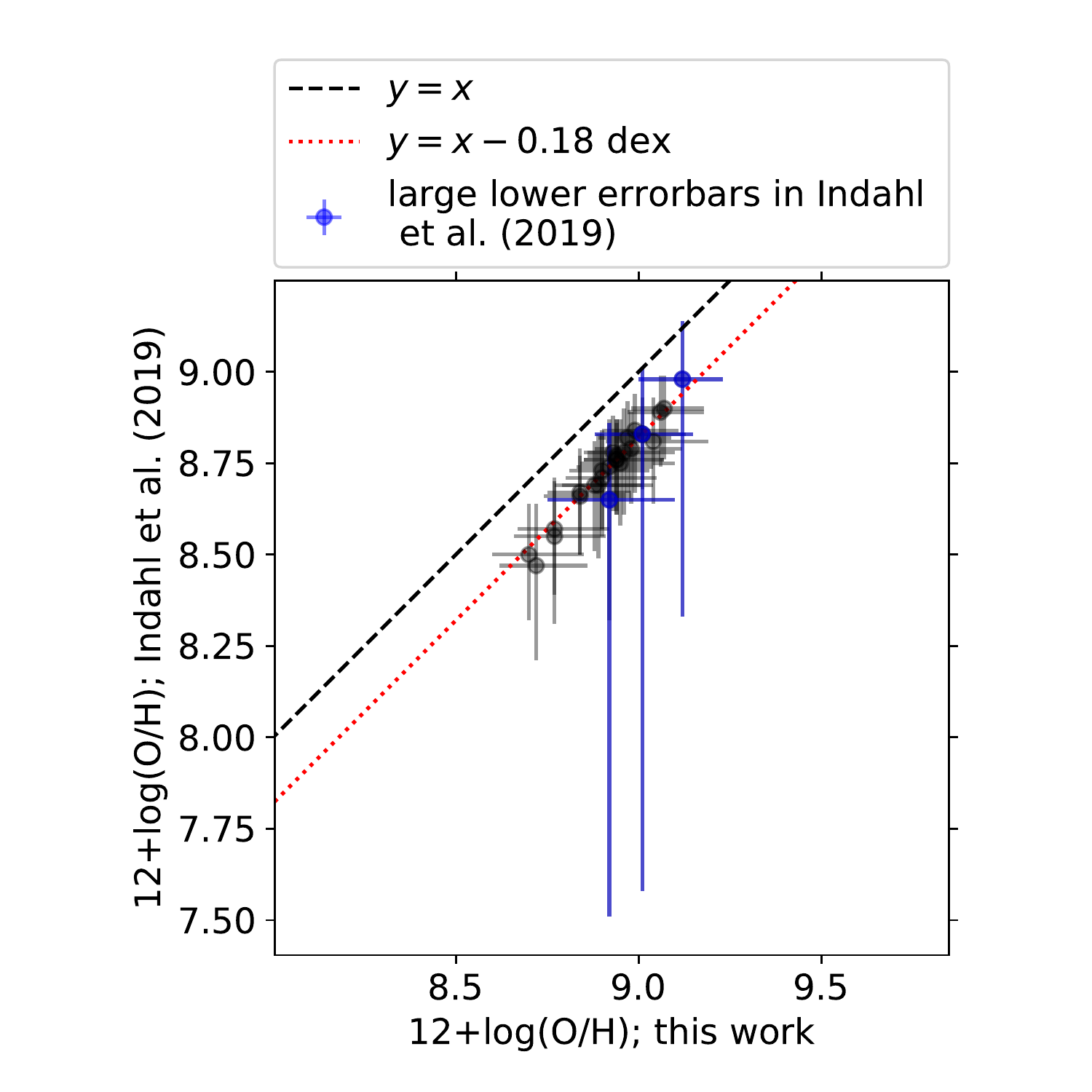}
    }
    \caption{Comparison between the metallicity estimates from this work and from \protect\cite{Indahl_2019_ApJ_883_114}. The black dashed and red dotted lines are the 1:1 \del{and 1:0.98 }correspondence \add{and $0.18$ dex shifted }lines\add{, respectively}. The three blue circles are the targets reported to have larger lower-credible-limits in \protect\cite{Indahl_2019_ApJ_883_114}: HPS067, HPS235, and HPS237.}
    \label{fig-metalcmp-median}
\end{figure*}

From here on, I compare my metallicity estimates, \metal, with those of \cite{Indahl_2019_ApJ_883_114}.
I skip the comparisons for \EBV{} and for \OIIIint{}, since their values are not reliable because of overestimation (see section \ref{ana-res-mock}).
In Fig.~\ref{fig-metalcmp-median}, I compare the metallicity estimates from this work with those from \cite{Indahl_2019_ApJ_883_114}.
Their median values are systematically lower than mine by \del{a factor of 0.98}\add{$0.18$ dex}.
The uncertainties are more-or-less similar to each other, except for the three targets that have much larger lower-credible-limits (HPS067, HPS235, and HPS237).
These targets have the three lowest S/Ns of [O II] and [O III] fluxes \citep[Table 3 of][]{Indahl_2019_ApJ_883_114} as mentioned above.
This low S/Ns may be related to the uncertainty differences for these three targets between this work and \cite{Indahl_2019_ApJ_883_114}.

\begin{figure*}
    \center{
    \includegraphics[scale=0.65]{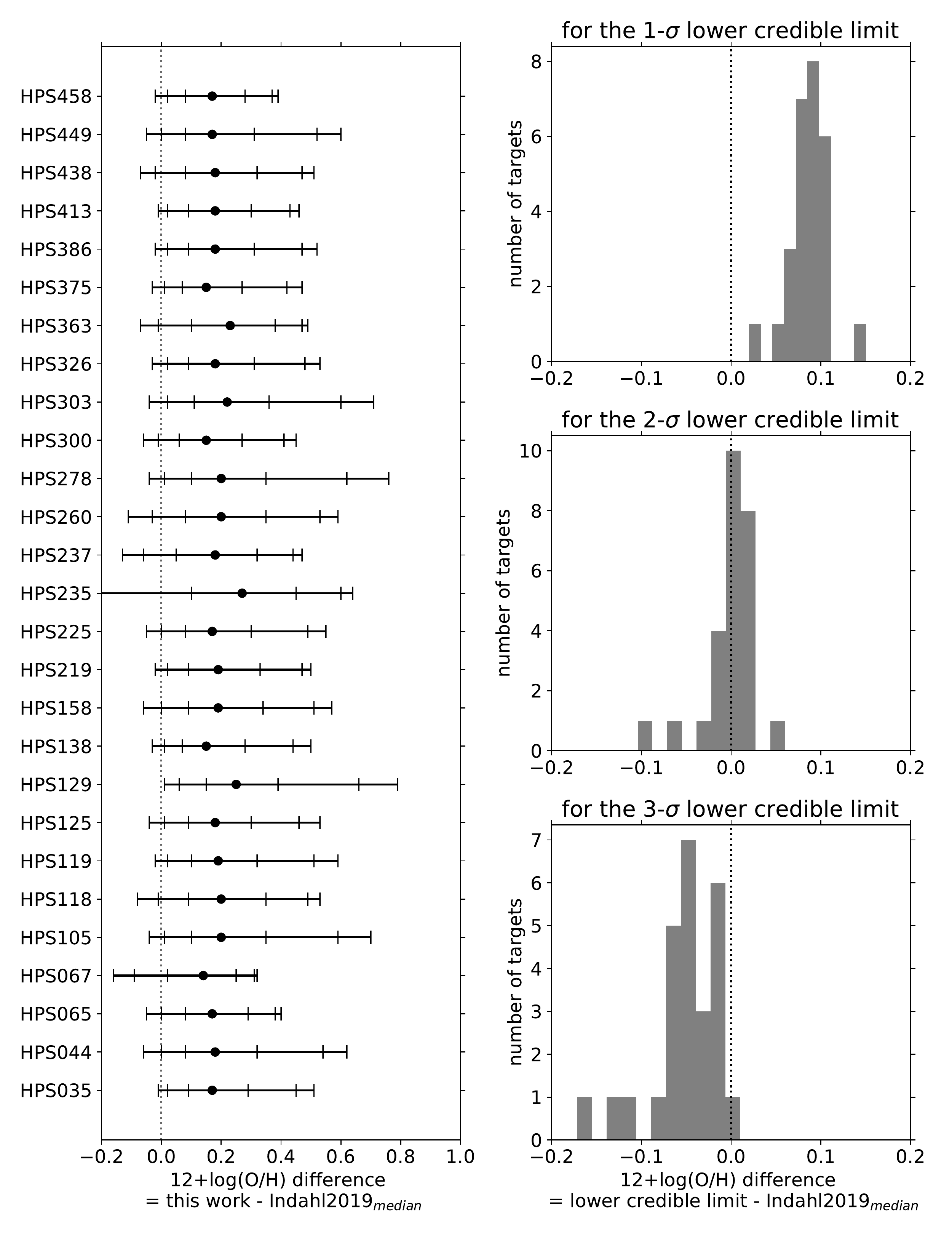}
    }
    \caption{Deviations between the metallicity estimates from this work and from \protect\cite{Indahl_2019_ApJ_883_114}. The dotted line in the left panel is the median metallicity estimated by \protect\cite{Indahl_2019_ApJ_883_114}, which is taken as the zero point for the present analysis. The circles with error bars in the left panel represent the metallicity distributions estimated in this work. The error bars are 1-$\sigma$ (68 \%), 2-$\sigma$ (95 \%), and 3-$\sigma$ (99 \%) credible intervals, respectively. The three right panels show how much the estimates from this work deviate from those of \protect\cite{Indahl_2019_ApJ_883_114}. Each histogram shows the 1-$\sigma$, 2-$\sigma$, or 3-$\sigma$ lower credible limit from this work minus the median values from \protect\cite{Indahl_2019_ApJ_883_114}. Again, the dotted lines in the right panels represent the median metallicity estimated by \protect\cite{Indahl_2019_ApJ_883_114}, which is taken as the zero point for the present analysis. The data point for HPS235 falls outside the frame in both the middle and bottom panels, because the 2-$\sigma$ and 3-$\sigma$ lower credible limit for HPS235 are much smaller than those for the other galaxies.}
    \label{fig-metalcmp-error}
\end{figure*}

Fig.~\ref{fig-metalcmp-error} shows how much the metallicity estimates from this work deviate from those of \cite{Indahl_2019_ApJ_883_114}.
The left panel of Fig.~\ref{fig-metalcmp-error} shows that all of my estimates are larger than theirs.
The right panels of Fig.~\ref{fig-metalcmp-error} show histograms of the difference between the lower credible limits from this work and the medians from \cite{Indahl_2019_ApJ_883_114}.
These histograms show that there is an overall 2-$\sigma$ difference between the metallicities from this work and from \cite{Indahl_2019_ApJ_883_114}. 

\begin{figure*}
    \center{
    \includegraphics[scale=0.65]{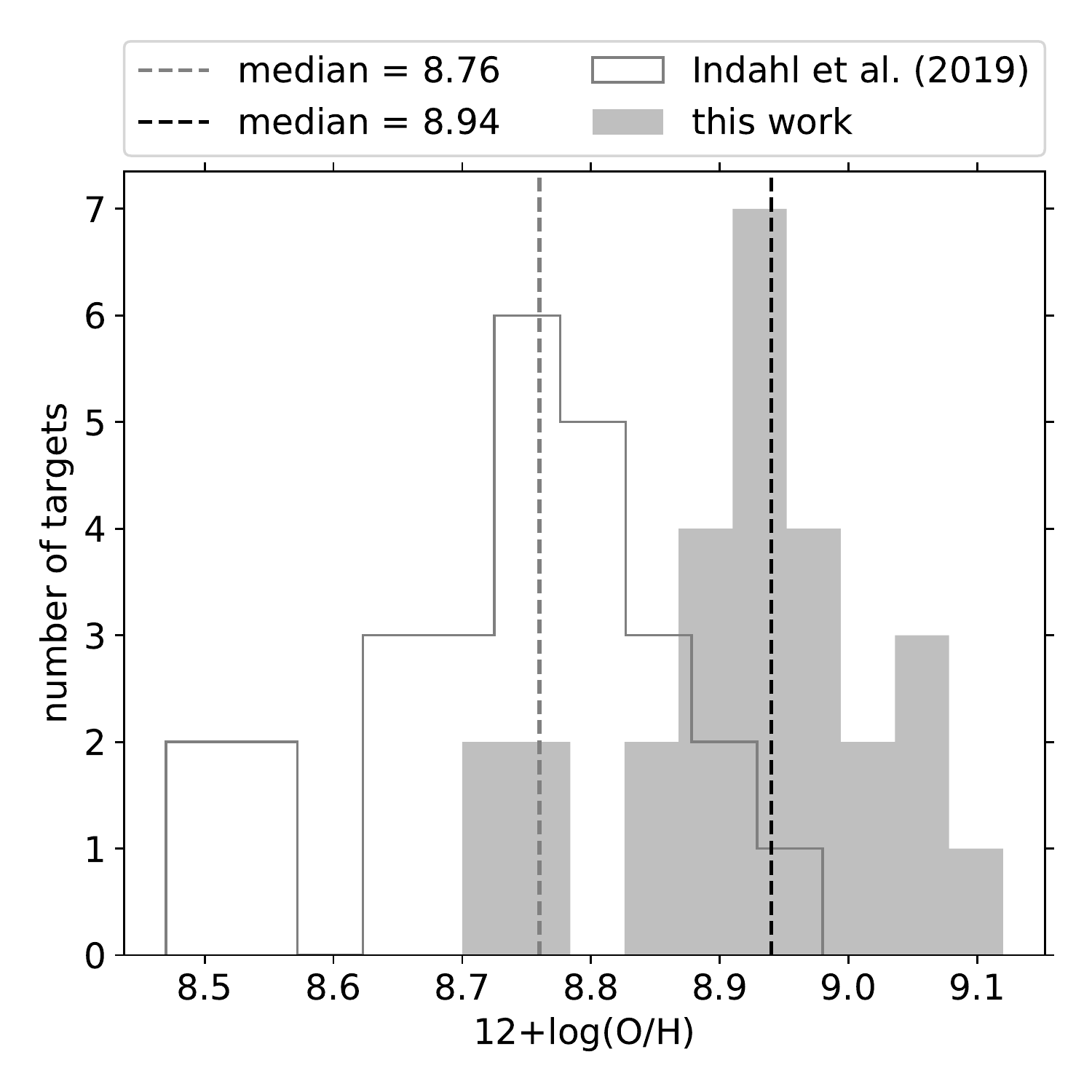}
    }
    \caption{Comparison of the metallicity distributions. The gray histograms are the median metallicity values from this work, and the white histograms are those from \protect\cite{Indahl_2019_ApJ_883_114}. The medians of the distributions are given in the legend.}
    \label{fig-metalcmp-dist}
\end{figure*}

Fig.~\ref{fig-metalcmp-dist} compares the two metallicity distributions.
Their shapes are similar, but my estimates are shifted to higher values, as expected from the systematically larger values of my estimates shown in Fig.~\ref{fig-metalcmp-median}.
The medians of the distributions are 8.94 for this work and 8.76 for \cite{Indahl_2019_ApJ_883_114}.

\begin{figure*}
    \center{
    \includegraphics[scale=0.65]{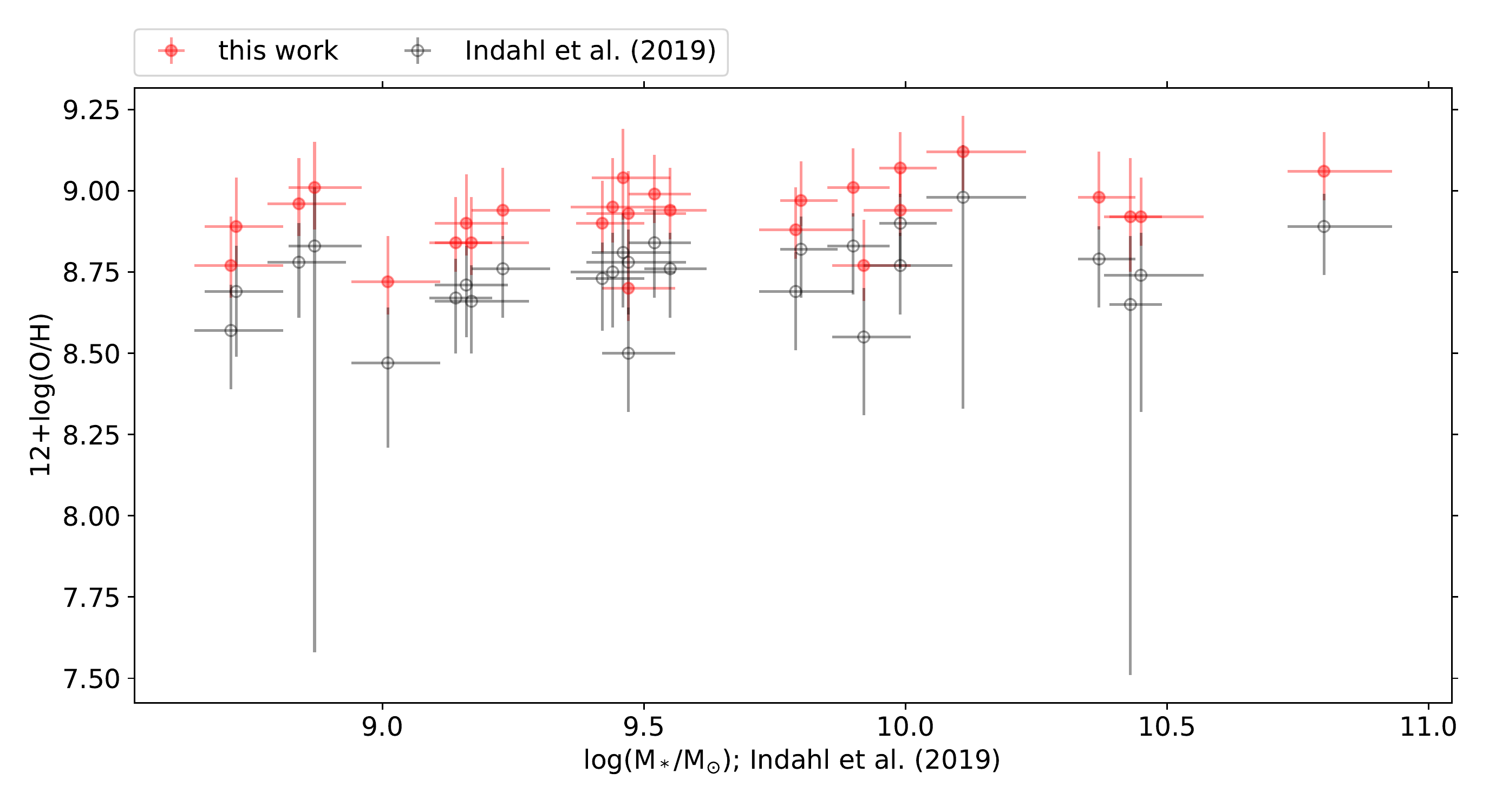}
    }
    \caption{Comparisons in the mass-metallicity plot. The red circles are the metallicity estimates from this work, and the gray circles are from \protect\cite{Indahl_2019_ApJ_883_114}. The mass estimates are from \protect\cite{Indahl_2019_ApJ_883_114}, which were derived from spectral-energy-distribution fitting with \texttt{MCSED} \citep{Bowman_2020_arXiv_06_13245}.}
    \label{fig-metal-mass}
\end{figure*}

Fig.~\ref{fig-metal-mass} compares the mass-metallicity plot obtained in this work with that from \cite{Indahl_2019_ApJ_883_114}.
The mass estimates are from \cite{Indahl_2019_ApJ_883_114}, which were derived from spectral-energy-distribution fitting with \texttt{MCSED} \citep{Bowman_2020_arXiv_06_13245}.
My metallicity estimates are a bit higher than those of \cite{Indahl_2019_ApJ_883_114}, although the 1-$\sigma$ uncertainties overlap each other somewhat.
The general trend of the almost-flat mass-metallicity relation is common to both estimates.

\section{Discussion} \label{disc}
The gas metallicities, \metal, I have newly determined are systematically higher than those of \cite{Indahl_2019_ApJ_883_114}, overall by about \del{a factor of $1/0.98\sim1.02$}\add{0.18 dex}, as shown in Fig.~\ref{fig-metalcmp-median}.
This trend is also shown in Figs.~\ref{fig-metalcmp-dist} and \ref{fig-metal-mass}.
Excluding three galaxies that have large lower-credible-limits (HPS067, HPS235, and HPS237) in the estimates from \cite{Indahl_2019_ApJ_883_114}, my 1-$\sigma$ uncertainty estimates are comparable to or smaller than those of \cite{Indahl_2019_ApJ_883_114} (Fig.~\ref{fig-metalcmp-median}).
These differences in estimates are probably due to the two factors that differ between \cite{Indahl_2019_ApJ_883_114} and this work---the uncertainties in the strong-line calibration (\del{Fig.~\ref{fig-model_test}}\add{see section \ref{ana-res-model}}) and the convergence monitoring of MCMC sampling---but it is hard to pinpoint the dominant one.
The three targets (HPS067, HPS235, and HPS237) have the three lowest S/Ns of [O II] and [O III] fluxes \citep[Table 3 of][]{Indahl_2019_ApJ_883_114}, but it is not clear whether this is the cause of the big differences in the uncertainty estimates between this work and \cite{Indahl_2019_ApJ_883_114}.
This big difference may be caused by a combination of the low S/Ns of the oxygen line fluxes and the different two factors in metallicity estimation mentioned above.

The \del{2 \%}\add{0.18 dex} difference in metallicity, \metal, mentioned above may seem insignificant, but I note that the median metallicities of \cite{Indahl_2019_ApJ_883_114} are smaller than mine by as much as 2-$\sigma$ in general (Fig.~\ref{fig-metalcmp-error}).
What is more important is the degree of convergence of the MCMC sampling.
I have shown that if one carries out MCMC sampling without monitoring the convergence, the ESS at the end of iteration can be as small as the one near the iteration start (compare Fig.~\ref{fig-hps-single} to Fig.~\ref{fig-hps}).
This means that the statistical accuracy of the MCMC sampling at the end of the iteration can be as poor as the one near the start of the MCMC sampling.
\delref{\cite{Indahl_2019_ApJ_883_114}}\del{ did not mention anything about the convergence of MCMC sampling.}
Therefore, it is improper to compare simply my results, which achieved ESS $>$ 2000, to \del{theirs}\add{\citeauthor{Indahl_2019_ApJ_883_114}'s} by assuming that both have the same statistical accuracy.

\cite{Indahl_2019_ApJ_883_114} compared their mass-metallicity relation to those of other galaxy populations, such as SDSS star-forming galaxies, and to some other extreme star-forming galaxies, like green-pea galaxies \citep{Cardamone_2009_MNRAS_399_1191}, blueberry galaxies \citep{Yang_2017_ApJ_847_38}, and blue compact dwarfs \citep{Sargent_1970_ApJ_162_L155,Kunth_2000_A&ARv_10_1}.
They estimated the metallicities of these reference galaxies, using their own strong-line method from the line fluxes in the literature: SDSS star-forming galaxies (median $z\sim0.078$, \citealt{Andrews_2013_ApJ_765_140}), green pea galaxies ($0.1\la z \la 0.4$, \citealt{Hawley_2012_PASP_124_21}), blueberry galaxies ($z \la 0.05$, \citealt{Yang_2017_ApJ_847_38}), and blue compact dwarfs ($0.2\la z \la0.5$, \citealt{Lian_2016_ApJ_819_73}).
Since (1) \cite{Indahl_2019_ApJ_883_114} used the same method to estimate the metallicities of both their target galaxies and the reference galaxies and (2) there is a systematic difference in metallicity between my estimates and those of \cite{Indahl_2019_ApJ_883_114} (Fig.~\ref{fig-metalcmp-median}), the relative difference in metallicity between the target galaxies of \cite{Indahl_2019_ApJ_883_114} and the reference galaxies would probably be the same, even if the metallicities of all the galaxies are estimated according to my procedure.
Therefore, the conclusion of \cite{Indahl_2019_ApJ_883_114} concerning the relative positions of the galaxies in the mass-metallicity plane would remain intact.
For example, the [O II]-selected galaxies from \cite{Indahl_2019_ApJ_883_114}, the SDSS star-forming galaxies, and the blue compact dwarfs would follow similar mass-metallicity relations, while the green-pea galaxies and blueberry galaxies would occupy a lower-metallicity region.
\add{Here I mention that using more recent and accurate calibration of \cite{Curti_2017_MNRAS_465_1384} in metallicity estimation would not make much difference in the mass-metallicity plot, because it returns a lower metallicity overall than the calibration of \cite{Maiolino_2008_A&A_488_463} as \cite{Indahl_2019_ApJ_883_114} showed in their Fig.~10. }
Additionally, I note that the location accuracy of a certain galaxy population in the mass-metallicity plane can be enhanced if the metallicity is estimated following my procedure, because I obtained much smaller credible limits for a few galaxies than did \cite{Indahl_2019_ApJ_883_114} (see Fig.~\ref{fig-metalcmp-median}).

As shown in section \ref{ana-res-mock}, the \EBV{} values estimated from this work and \cite{Indahl_2019_ApJ_883_114} are not reliable, because they are overestimated (see Figs.~\ref{fig-mock-EBVpriorG}-\ref{fig-mock-metal9}).
\cite{Indahl_2019_ApJ_883_114} used the estimated \EBV{} to correct the [O II] line fluxes, and this consequently affects the SFR estimates.
A typical value of the \EBV{} obtained both in this work and in \cite{Indahl_2019_ApJ_883_114} is about $0.40\pm0.15$.
If \EBV{} is overestimated by about 2-$\sigma$, as shown in Figs.~\ref{fig-mock-EBVpriorG}-\ref{fig-mock-metal9}, the true \EBV{} would be around $0.4-0.15\times2=0.1$.
I calculated the overestimation factor for the [O II] $\lambda3727$ emission line using the Calzetti attenuation curve \citep{Calzetti_2000_ApJ_533_682}, and found it to be $8.65/1.72\sim5.05\sim0.70$ dex.
This means that the SFR estimates of \cite{Indahl_2019_ApJ_883_114} can be overestimated by a factor of five, and thus some of their conclusions regarding the SFR \del{should be revised}\add{need to be reconsidered}.

First, \cite{Indahl_2019_ApJ_883_114} found that the SFR distribution of their target galaxies was similar to those of the SDSS star-forming galaxies \citep{Andrews_2013_ApJ_765_140}.
\del{This statement should be revised as follows: the SFR distribution of }\delref{\cite{Indahl_2019_ApJ_883_114}}\del{ is probably lower in actuality than that of the SDSS star-forming galaxies.}\add{With our analysis, the SFR distribution of \citeauthor{Indahl_2019_ApJ_883_114}'s galaxies turns out to be lower than that of the SDSS star-forming galaxies.}
The difference may be smaller because the SFR of the SDSS star-forming galaxies is the total galactic SFR, instead of being calculated only from the light within the fiber as in \cite{Indahl_2019_ApJ_883_114}. 

Second, \cite{Indahl_2019_ApJ_883_114} found that their [O II]-selected galaxies reside between the green-pea galaxies \citep{Hawley_2012_PASP_124_21} and the SDSS star-forming main-sequence galaxies \citep{DuartePuertas_2017_A&A_599_A71} along the SFR axis in the mass-SFR plane, implying that their blind spectroscopic survey fills in a galaxy population that SDSS missed.
Since the SFRs of the green-pea galaxies and of \citeauthor{Indahl_2019_ApJ_883_114}'s [O II]-selected galaxies were determined using the overestimated \EBV{} from \cite{Indahl_2019_ApJ_883_114}, the SFRs of these galaxies should be lower than the values presented in \cite{Indahl_2019_ApJ_883_114}.
Therefore, the SFRs of the [O II]-selected galaxies are not likely to exceed the values of the star-forming main-sequence galaxies, which undermines their conclusion about filling in a missed galaxy population.
However, \cite{Indahl_2019_ApJ_883_114} have only 27 [O II]-selected galaxies.
Considering the 680-times-larger survey volume and the lower line-flux limit of the HETDEX than of the HPS \citep{Indahl_2019_ApJ_883_114}, it is premature to conclude that such a missed galaxy population would not be discovered from the forthcoming HETDEX survey.

\begin{figure*}
    \center{
    \includegraphics[scale=0.65]{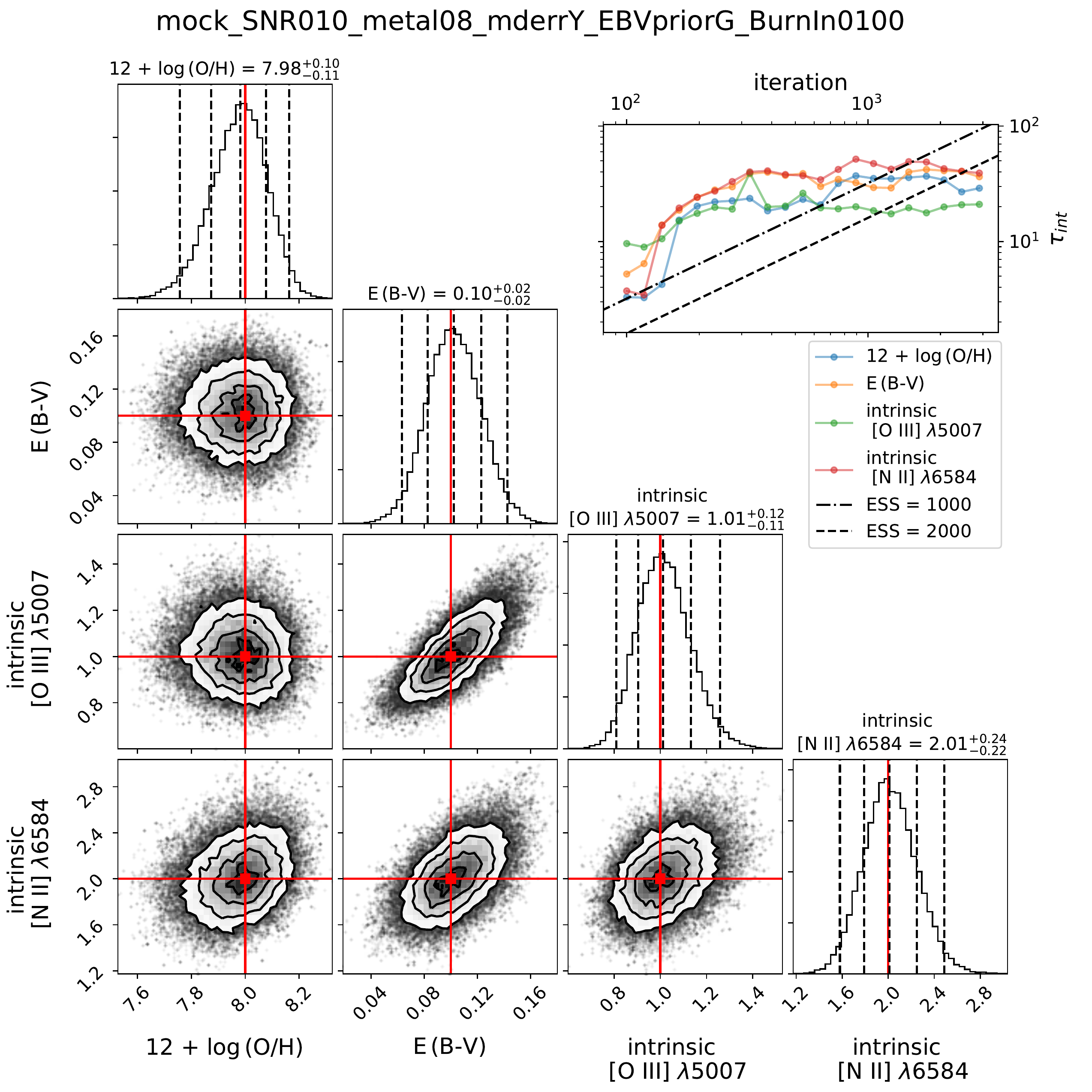}
    }
    \caption{Posterior distribution and evolution of the integrated autocorrelation times (\tauint) from another mock-data test. The setting for this test is the same as for Fig.~\ref{fig-mock-EBVpriorG}, except that I have adopted a narrower Gaussian prior for \EBV{} that is centered at the input value (0.1) with $\sigma=0.02$ in order to mimic an independently determined \EBV{}. The figure description is otherwise the same as for Fig.~\ref{fig-mock-snr100}.}
    \label{fig-mock-EBVnarrow}
\end{figure*}

As shown in section \ref{ana-res-mock}, the model parameter \EBV{} does not recover the input value well.
Therefore, \EBV{} should be determined another way in order to correct the reddening properly.
Since the HETDEX does not cover H$\alpha$, one simple way would be to use the Balmer decrement between H$\beta$ and H$\gamma$.
I have verified that when \EBV{} is correctly and independently estimated, the other three model parameters---\metal{}, \OIIIint{}, and \NIIint{}---recover the input values to within the 1-$\sigma$ level (Fig.~\ref{fig-mock-EBVnarrow}).
In Fig.~\ref{fig-mock-EBVnarrow}, I have adopted a narrow Gaussian prior for \EBV{} that is centered at the input value (i.e., 0.1) with $\sigma=0.02$ in order to mimic the independently determined \EBV{}.
I also checked how many targets would show a detectable \Hg{} line in the HETDEX survey.
The HETDEX survey has a sensitivity of $\sim3.5\times 10^{-17}$ \flux{} in a baseline 20 min observation \citep{Hill_2016_inproc}.
Assuming \EBV{} = 0.3 (close to the mean of the prior), Calzetti's attenuation curve \citep{Calzetti_2000_ApJ_533_682}, and Case B recombination of Balmer lines (flux ratio of H$_\gamma$ to H$_\beta\sim0.47$, \citealt{Osterbrock_2005_book}), I found that the reddened \Hb{} flux of $\sim1\times10^{-16}$ \flux{} corresponds to the reddened \Hg{} flux similar to the HETDEX sensitivity.
If I apply this \Hb{} cutoff value to the galaxies observed in \cite{Indahl_2019_ApJ_883_114}, 20 out of 29 galaxies would show a detectable \Hg{} line in the HETDEX survey.
Not to lose the rest 1/3 of the targets, the attenuation correction should be done using \Ha{} and \Hb{} lines instead, which consequently demands the follow-up observations.
Another way to avoid the \EBV{} overestimation is to use another strong-line calibration that shows much less scatter of line ratios for a given metallicity.
However, comparable scatters are seen in other more recent calibrations that use the electron-temperature method \citep[e.g.,][]{Pilyugin_2016_MNRAS_457_3678,Curti_2017_MNRAS_465_1384}.
Thus, this way does not seem to alleviate the \EBV{} overestimation problem.

\section{Conclusion}
I have reanalyzed the local ($z<0.15$) star-forming galaxies of \cite{Indahl_2019_ApJ_883_114} and have newly determined the gas metallicities, \metal.
\citeauthor{Indahl_2019_ApJ_883_114}'s target galaxies are from the HPS, which is a pilot survey for the more extensive, IFU-based, blind spectroscopic survey HETDEX.
The HPS covers $\sim3500-5800$ \AA{} at $\sim5$ \AA{} resolution.
\cite{Indahl_2019_ApJ_883_114} collected the line-emitting galaxies from this survey and measured the line fluxes of [O II] $\lambda$3727, H$\beta$, [O III] $\lambda$5007, etc.
They estimated the gas metallicities of the galaxies using the strong-line method, which employs the flux ratios of strong (i.e., easy to observe) emission lines, employing the Bayesian approach and MCMC sampling.
However, I noticed \del{that}\add{three points that can be improved in their analysis:} (1) they had adopted a relatively small uncertainty for the line-ratio calibration; (2) they had not \del{carried out}\add{presented} reproducibility tests with mock data; and (3) they had not \del{monitored}\add{mentioned} the convergence of the MCMC sampling, which is important for ensuring the statistical accuracy of the MCMC samples.
Here I reanalyzed the 27 [O II]-selected galaxies from \cite{Indahl_2019_ApJ_883_114}, following their analysis scheme but carefully dealing with the three points mentioned above.
I adopted a higher uncertainty for the line-ratio calibration to mimic properly that of the calibration \cite{Indahl_2019_ApJ_883_114} had adopted \citep{Maiolino_2008_A&A_488_463}, and I secured ESS $>$ 2000 for all the MCMC samples\add{{ }to achieve enough convergence}.
From reproducibility tests with mock data, I found that the parameter-estimation method of \cite{Indahl_2019_ApJ_883_114} can overestimate \EBV{}, \OIIIint{}, and \NIIint{} by as much as 2-$\sigma$ or more, although it can recover the metallicity to within 1-$\sigma$.
Therefore, among the four model parameters only the metallicity estimates are reliable.
When reanalyzing the HPS data, I excluded two [O III]-selected galaxies \citep[see][]{Indahl_2019_ApJ_883_114} because of the difficulty in one-to-one comparison between this work and \citeauthor{Indahl_2019_ApJ_883_114}'s.
Therefore, I only used the ratios R23 [eq.~(\ref{R23})] and O32 [eq.~(\ref{O32})], and hence needed three model parameters: \metal{}, \EBV, and \OIIIint{} (see section \ref{ana-res-hps}).

My metallicity determinations, \metal, are systematically higher than those of \cite{Indahl_2019_ApJ_883_114} by \del{a factor of $1/0.98\sim1.02$}\add{0.18 dex} (Fig.~\ref{fig-metalcmp-median}).
Although this factor may seem insignificant, the median metallicities of \cite{Indahl_2019_ApJ_883_114} deviate from my estimates by as much as 2-$\sigma$ overall (Fig.~\ref{fig-metalcmp-error}).
The more important issue, however, concerns the convergence of the MCMC sampling.
\cite{Indahl_2019_ApJ_883_114} did not \del{monitor}\add{mention} the convergence, so one cannot know whether or not their estimates have appropriate statistical accuracy.
I have shown that the statistical accuracy of the MCMC sampling can be as poor as the one near the iteration start if one runs the MCMC sampling without monitoring the convergence (see Fig.~\ref{fig-hps-single} and its counterpart Fig.~\ref{fig-hps}).
However, the conclusion of \cite{Indahl_2019_ApJ_883_114} about the relative position of their [O II]-selected galaxies to other star-forming galaxy populations in the mass-metallicity plane would remain intact, because (1) \cite{Indahl_2019_ApJ_883_114} used the same method to estimate the metallicity for all the galaxies under comparison and (2) there is a systematic difference between my estimates and those of \cite{Indahl_2019_ApJ_883_114} (Fig.~\ref{fig-metalcmp-median}).

Considering the \EBV{} overestimation, \add{we need to rethink }\citeauthor{Indahl_2019_ApJ_883_114}'s two conclusions about the SFR\del{ should be revised}, because they corrected the reddening using \EBV{}, which can overestimate the SFR by as much as a factor of five (see section \ref{disc}).
First, \citeauthor{Indahl_2019_ApJ_883_114}'s galaxies probably have a `lower' SFR distribution than the SDSS star-forming galaxies \citep{Andrews_2013_ApJ_765_140}, instead of a similar SFR distribution to the SDSS one as \cite{Indahl_2019_ApJ_883_114} described.
Second, \cite{Indahl_2019_ApJ_883_114} found that their [O II]-selected galaxies reside between the green-pea galaxies (a kind of extreme star-forming galaxies) and the SDSS star-forming main-sequence galaxies along the SFR axis in the mass-SFR plane.
In actuality, this is probably not the case, because the SFRs of \citeauthor{Indahl_2019_ApJ_883_114}'s galaxies and of the green-pea galaxies have been overestimated.
That is, \citeauthor{Indahl_2019_ApJ_883_114}' galaxies are likely to overlap with the main-sequence galaxies.
This undermines \citeauthor{Indahl_2019_ApJ_883_114}'s conclusion that their blind spectroscopic survey fills in a galaxy population that has been missed in photometric surveys with continuum-flux preselection, like the SDSS.
However, it is premature to conclude that the forthcoming HETDEX would not discover hitherto-unknown galaxy populations, considering that the HETDEX will have a much larger survey volume and lower line-flux limit than the HPS.
To avoid the \EBV{} overestimation, I checked whether the independent determination of \EBV{} from H$\beta$ and H$\gamma$ lines is a viable option\del{ for the correct parameter estimation}. 
I found that once \EBV{} is independently determined, the other three model parameters---\metal{}, \OIIIint, and \NIIint---recover the input values well (Fig.~\ref{fig-mock-EBVnarrow}).
I also found that $\sim$2/3 of the galaxies observed in \cite{Indahl_2019_ApJ_883_114} would show a detectable \Hg{} line in the HETDEX survey, although another follow-up observation would be required not to lose the rest 1/3 of the galaxies (for the independent \EBV{} determination using \Ha{} and \Hb{} lines).

\section*{Acknowledgements}
\add{J.-H.S. is grateful to the referee for his useful and considerate comments, which improved the manuscript and will help the author in his future endeavors.}
J.-H.S. is\add{{ }also} thankful to Ho Seong Hwang for several useful discussions.

\section*{Data Availability}

All the MCMC sampling results can be downloaded from \del{\url{http://vo.kasi.re.kr/Stat_Reanal/}}\add{\url{https://data.kasi.re.kr/vo/Stat_Reanal/}} with a plotting script.



\bibliographystyle{mnras}
\bibliography{stat_reanal_2019a} 








\bsp	
\label{lastpage}
\end{document}